\documentclass[acmsmall]{acmart}

\AtBeginDocument{%
  \providecommand\BibTeX{{%
    \normalfont B\kern-0.5em{\scshape i\kern-0.25em b}\kern-0.8em\TeX}}}

\setcopyright{acmcopyright}
\copyrightyear{2020}
\acmYear{2020}
\acmDOI{10.1145/1122445.1122456}

\acmJournal{JACM}
\acmVolume{37}
\acmNumber{4}
\acmArticle{111}
\acmMonth{8}

\newtheorem{definition}{Definition}

\usepackage{pdflscape}
\usepackage{enumitem}
\usepackage{multirow}

\begin{document}

\title{Narrative Maps: An Algorithmic Approach to Represent and Extract Information Narratives}

 \author{Brian Keith Norambuena}
 \affiliation{%
   \institution{Virginia Tech}
   \department{Department of Computer Science}
   \streetaddress{620 Drillfield Drive}
   \city{Blacksburg}
   \state{VA}
   \country{USA}
   \postcode{24061}
 }
  \additionalaffiliation{%
   \institution{Universidad Católica del Norte}
   \department{Department of Computing \& Systems Engineering}
   \city{Antofagasta}
   \country{Chile}
   \postcode{1270709}
  }
 \email{briankeithn@vt.edu}
 \author{Tanushree Mitra}
 \affiliation{%
   \institution{University of Washington}
   \department{The Information School}
   \city{Seattle}
   \state{WA}
   \country{USA}
 }
 \email{tmitra@uw.edu}

\renewcommand{\shortauthors}{Keith and Mitra}
\mathchardef\mhyphen="2D

\begin{abstract}
Narratives are fundamental to our perception of the world and are pervasive in all activities that involve the representation of events in time. 
Yet, modern online information systems do not incorporate narratives in their representation of events occurring over time. This article aims to bridge this gap, combining the theory of narrative representations with the data from modern online systems. We make three key contributions: a theory-driven \emph{computational representation} of narratives, a novel \emph{extraction algorithm} to obtain these representations from data, and an \emph{evaluation} of our approach. In particular, given the effectiveness of visual metaphors, we employ a route map metaphor to design a narrative map representation. The narrative map representation illustrates the events and stories in the narrative as a series of landmarks and routes on the map. Each element of our representation is backed by a corresponding element from formal narrative theory, thus providing a solid theoretical background to our method. Our approach extracts the underlying graph structure of the narrative map using a novel optimization technique focused on maximizing coherence while respecting structural and coverage constraints. We showcase the effectiveness of our approach by performing a user evaluation to assess the quality of the representation, metaphor, and visualization. Evaluation results indicate that the Narrative Map representation is a powerful method to communicate complex narratives to individuals. Our findings have implications for intelligence analysts, computational journalists, and misinformation researchers.
\end{abstract}



\maketitle

\section{Introduction}
\begin{quote}
\textit{People don't see the world before their eyes until it's put in a narrative mode.}\begin{flushright}\indent\indent- Filmmaker Brian De Palma \cite{abbott2008cambridge}\end{flushright}
\end{quote}

Narratives are fundamental to our understanding of the world \cite{abbott2008cambridge} and are central to human relations \cite{miskimmon2014strategic}. They are the frameworks that enable humans to associate otherwise unconnected events \cite{burke1969grammar} and play a key role in collaborative sensemaking in society \cite{baber2011sensemaking, wilson2018assembling}. For decades, communication science scholars have emphasized the importance of narratives in human communication \cite{abbott2008cambridge}. Narrative theorists have gone so far to claim that ``narrative is the principled way in which our species organizes its understanding of time.'' 

In this line of research, narratives are defined as a \emph{coherent system of interrelated stories} \cite{halverson2011master} and are considered to be omnipresent in all activities that involve the representation of events in time. Through these systems of stories, humans produce a shared understanding of the world \cite{szostek2017defence}. 

Despite the  compelling case of narratives in human communication, modern-day online information systems do not incorporate narratives in their representation of events happening over time. Instead, most representations are still chronological order of events from the earliest to the latest; for example, consider Google News or Twitter's news feed.

This paper aims to bridge this gap, merging the theory behind narrative representations with data left in modern online systems (specifically news systems) while leveraging graph-driven approaches to identify such narrative structures. With theory as a guide, our central question is: Can we computationally represent and extract narratives from large-scale online data? 

Our contribution is not just methodological; narrative representation has the potential to significantly impact online information consumers. Consider the unrelenting barrage of online news articles that users get exposed to or the ever-increasing problem of information overload in online systems \cite{ho2001towards,shahaf2010connecting}. Not just regular users, even expert news investigators find it daunting to keep track of the evolving storylines, often losing track of the big picture. Given narratives are the primary mode in which humans apprehend time-based information \cite{abbott2008cambridge}, our paper offers just such a representation and the algorithmic approach to surface them directly from data. 

In particular, we use a route map metaphor to provide an intuitive basis for our representation. This metaphor helps connect the elements of narrative theory and graph theory. Building upon this representation (RQ1), we propose a novel extraction algorithm to generate a narrative map representation automatically from data (RQ2). Our algorithmic approach leverages optimization techniques and graph-theoretical concepts to find all the elements in our narrative representation.

Specifically, our work addresses three research questions. First, we formalize the computational representation of narratives and ask:

\begin{quote}
\textbf{RQ1 Representation}: How can we computationally represent a narrative?
\end{quote}

Based on a route map metaphor, we define a computational representation of a narrative grounded in formal narrative theory. Our representation maps  concepts from narrative theory into graph theory elements. In particular, our proposal considers the use of a directed acyclic graph since this structure naturally reflects the temporal structure of stories. We also consider edge weights that allow us to evaluate the coherence of the storylines in the narrative map. Note that a narrative is composed of at least one storyline. After identifying the elements of our representation, we focus on how to extract these elements from data.

\begin{quote}
\indent\indent \textbf{RQ2 Extraction}: How can we extract a narrative representation from data?
\end{quote}

To extract the narrative representation, we design an optimization approach that seeks to maximize the coherence of the narrative map subject to structural and topic coverage constraints. After generating the basic structure of the narrative map, we extract the main route (the most coherent path in the narrative) and the representative landmarks (sets of representative events from each alternative storyline). Using this information we can infer the themes---the recurrent or prominent ideas---of the narrative. Finally, we evaluate how well our proposed approach works.

\begin{quote}
\indent\indent \textbf{RQ3 Evaluation}: How well does our approach work to represent narratives?
\end{quote}

We present a detailed evaluation of our proposed approach. In particular, we study the news narrative of the Coronavirus outbreak with three maps from the first three months of 2020, each generated after processing hundreds of news articles.  We find that our representation provides a coherent narrative and that our metaphors are appropriate to model the narrative. We also show how our representation provides an easy way for users to answer questions such as: What is the main storyline---the main route---of this narrative? Apart from the main storyline, what other alternative storylines are there in the narrative? What are the major themes in the narrative?

Thus, by answering these research questions, our key contributions are: 
\begin{itemize}
\item A \textbf{computational representation} for narratives, based on a mapping between narrative theory concepts and computational elements grounded in the route map metaphor. Our representation also includes a visualization based on the route map metaphor. 
\item An \textbf{extraction algorithm} to construct a narrative representation from data while using an optimization method and graph-theoretical concepts.
\item An \textbf{evaluation} of our approach through a case study and a user study. In particular, we test our methods on the news narrative of the first three months of the Coronavirus outbreak.
\end{itemize}

Furthermore, we note that our work builds upon the foundation created by existing techniques. However, our proposed representation extends them by providing a theoretical framework based on formal narrative theory and the route map metaphor. The inclusion of this framework distinguishes our work from similar methods.

We believe that our approach has widespread implications. At its core, our narrative representation and our extraction method can help users understand the big picture of a narrative. Moreover, expert users, such as journalists, fact-checkers, and intelligence analysts can leverage our extraction method to explore the narrative landscape in the context of their domain of work. Furthermore, our narrative representation could be enhanced by considering additional elements, such as contextual information, credibility and political bias annotations, or more detailed event descriptors. In particular, adding credibility and political bias information can help understand how misinformation spreads in news narratives.

In the rest of this section, we introduce the route map metaphor and provide a motivating example, explaining the reasoning behind the metaphor and detailing how it can be applied to analyze a narrative. In Section 2, we introduce important background concepts and describe related works. In Section 3, we show our algorithmic approach, including both the computational representation of narratives and the extraction algorithm itself. In Section 4, we present a case study of the Coronavirus narrative. Section 5 presents a user evaluation of our approach followed by a discussion of our results in Section 6. Finally, we end with the main conclusions of our work.

\subsection{Route Map Metaphor}
Humanity has relied on maps and cartography to understand its physical surroundings since antiquity. In particular, route maps and similar cartographic representations have been used to describe routes, that is, how to get from one physical place to another. These routes are usually accompanied by information of places and important locations along the way, as well as additional information about relatively less important elements of its surroundings. There can also be many routes joining two places, with differing characteristics and importance. 

In general, the idea of using maps to represent stories is called information cartography \cite{shahaf2013information}. In this paper, we propose a way to represent a narrative (i.e., a system of stories) using a narrative route map (or simply narrative map). Instead of joining two physical places, we join two narrative places (a starting and ending event). The locations shown along the routes in our map should correspond to narrative elements.

\begin{figure}
    \centering
    \includegraphics[width=\textwidth]{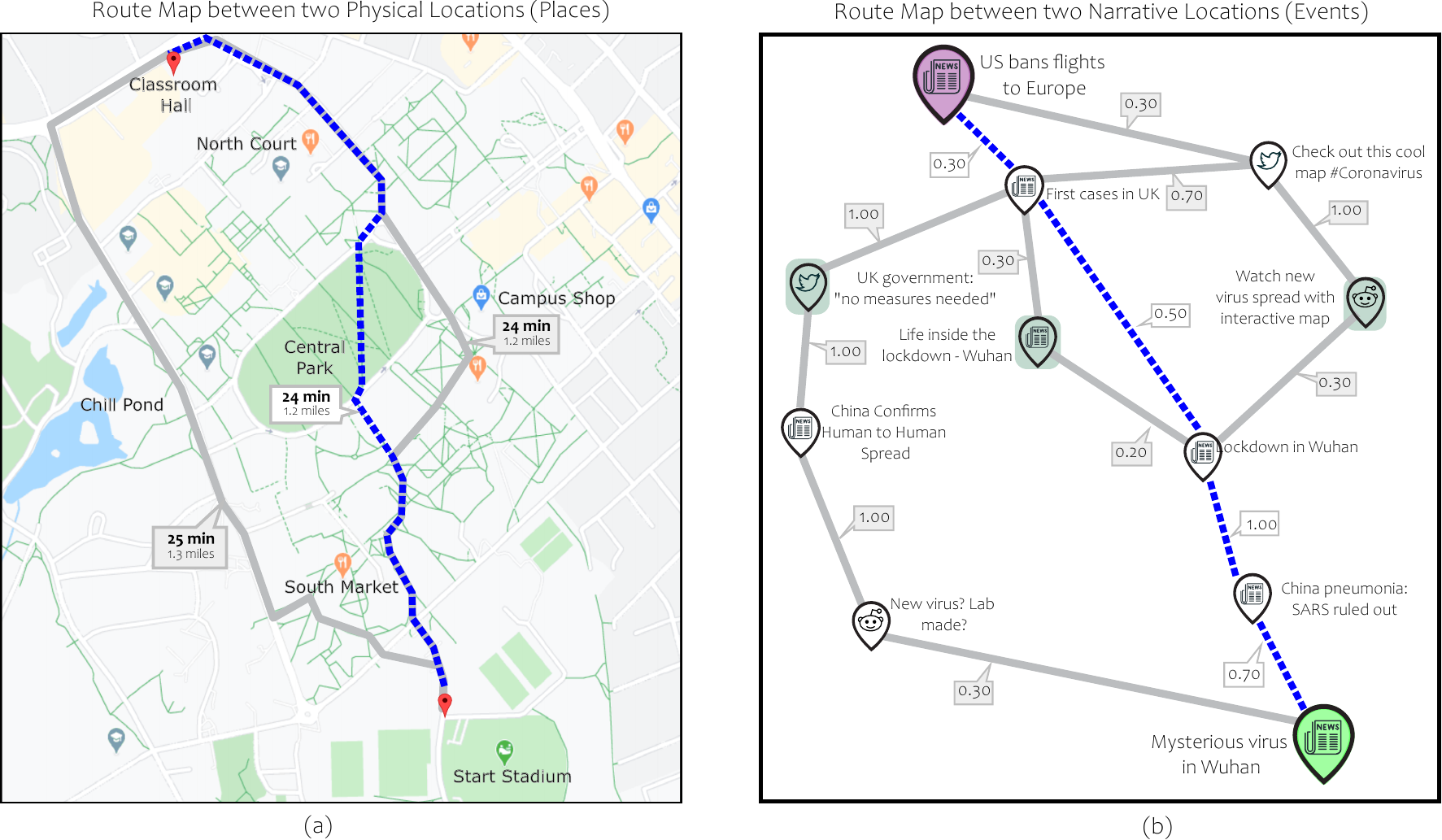}
    \caption{Illustration of the Route Map Metaphor. (a) A route map showing how to get from Start Stadium to Classroom Hall. (b) A narrative route map showing how to get from the Mysterious Virus in Wuhan event (starting event \protect\resizebox{0.30cm}{!}{\protect\includegraphics{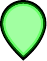}} ) to the US banning flights to Europe (ending event \protect\resizebox{0.30cm}{!}{\protect\includegraphics{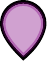}} ). We highlight some representative landmarks \protect\resizebox{0.35cm}{!}{\protect\includegraphics{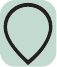}} for each route in the narrative map.} 
    \label{fig:metaphor}
\end{figure}

Consider a route map that shows routes between two physical places as shown in Figure \ref{fig:metaphor}(a). The routes connect the ``Start Stadium'' and the ``Classroom Hall'' landmarks. Suppose a student needs to go to his engineering class at Classroom Hall after watching a practice session at Start Stadium. However, they are not sure which route to take, so they would take out their phone, enter their current location, and where they want to go into the route map application. The application then provides them with a route map as shown in Figure \ref{fig:metaphor}. We can identify multiple elements of interest in this map. In particular, it contains multiple interconnected \textit{routes} between the \textit{starting place} (Start Stadium) and the \textit{ending place} (Classroom Hall).

One of these routes is highlighted as the \textit{main route} (in blue) because it is the shortest path. All routes contain information about its distance and the time it takes to go through them, this allows users to decide which one to take (i.e., it is an \textit{evaluation metric}), although in this particular example there is not much difference between the three routes. 

We can also see information about different kinds of \textit{landmarks} throughout each route. We note that there are several types of landmarks (e.g., restaurants, shops, parks, and educational buildings). For example, the main route passes close to the South Market and goes through the Central Park, and then North Court before getting to the Classroom Hall. 

In fact, we could identify each route by mentioning a \textit{representative landmark}, such as the Chill Pond, the Central Park, and the Campus Shop for each route in our case. By identifying these representative landmarks, we can differentiate among these routes. Someone using the route map could make choices informed by these representative landmarks. For example, someone seeking to walk through a more scenic view might choose to go through the Chill Pond route, while someone interested in buying a snack or some materials could go through the Campus Shop route. Even though neither of these routes is the main one they still offer a potentially useful alternative.

From the previous analysis we identify the following key elements of a route map: a starting place, an ending place, routes joining these places, landmarks along those routes, representative landmarks for parallel routes, an evaluation metric (e.g., distance or time) for those routes, and the main route obtained using that metric. Using these elements, we construct a metaphorical representation of narratives with a route map. Just as physical landscapes can be represented using these route maps, we use the same elements to construct a route map representation of the narrative landscape surrounding a certain event. 

According to some authors \cite{abbott2008cambridge}, narratives must contain at least two events: a starting event \resizebox{0.30cm}{!}{\includegraphics{figures/start.pdf}} and ending event \resizebox{0.30cm}{!}{\includegraphics{figures/end.pdf}}. These events work similarly to the starting and ending place on the map and we need to choose them before creating the narrative map. We are interested in understanding how these two events are joined together by a story (i.e., a series of events and discourse elements). Thus, events would be akin to the landmarks of each route \resizebox{0.30cm}{!}{\includegraphics{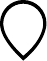}}, and discourse elements would be the additional information associated with them. Furthermore, narratives can be seen as a system of stories, which in our metaphor correspond to the routes on the map \resizebox{0.35cm}{!}{\includegraphics{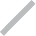}}. By doing this we gather new insights about the original events and the surrounding themes of the underlying stories. In addition, we can identify and characterize these storylines by finding a representative landmark for them \resizebox{0.35cm}{!}{\includegraphics{figures/representative_landmarks.pdf}}. Finally, we can evaluate the stories \resizebox{0.35cm}{!}{\includegraphics{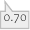}} (e.g., by looking at its coherence) to find the main storyline of the narrative \resizebox{0.35cm}{!}{\includegraphics{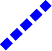}} (i.e., the main route).

See for example Figure \ref{fig:metaphor}(b), which shows a metaphoric rendition of a narrative route map representing the first month of the Coronavirus outbreak timeline. This map contains several sources of information (news and social media) and provides an overview of the evolving narrative landscape during this period. The dashed line \resizebox{0.35cm}{!}{\includegraphics{figures/main_route.pdf}} represents the main route (i.e., storyline). It reflects the development of the outbreak from its first cases as a mysterious virus to the US banning flights. 

Various other elements are intertwined with this route, representing alternative paths towards the ending event. Representative landmarks \resizebox{0.35cm}{!}{\includegraphics{figures/representative_landmarks.pdf}}, in this case, could be the tweets about the UK government indicating that no measures are needed, the news article about the life inside the lockdown in Wuhan, and the submission on Reddit about watching how the virus spreads. 

As in the physical route map, these landmarks can inform the user of the map on where to go. For example, a data analyst could be interested in reading about visualizations related to the virus as examples of data visualizations. This analyst would presumably want to go through the interactive map route. A social science researcher might be interested in looking at how news sources presented life during lockdown at the start of the crisis. Therefore, they would likely be interested in reading the lockdown route. Finally, a political activist from the UK could be interested in reading about the UK response to the virus to build a case against the government's handling of the outbreak. Thus, they would probably find the route on the UK response more interesting.

In one of the paths we see conspiracy talks about the virus being artificial, then news about human to human spreading, leading into the spread into other countries. In the continuation of these paths, we see how the lack of appropriate government measures leads to the spread of the virus. Some paths are simply smaller deviations with interesting tidbits of information, such as the article on life during lockdown and people sharing an interactive map in social media.

This route map metaphor forms the intuitive basis of our computational representation and corresponding visualization. However, to give our intuition a stronger foundation, we associate each of the elements of our representation with a formal concept from narrative theory.

\section{Background and Related Work}
In this section, we introduce important background concepts from narrative theory and present research related to our main contributions. We start with online narratives, then we move onto narrative theory definitions, and finally, we describe existing approaches to extract and represent narratives and stories.

\subsection{Online Narratives}
The exponential growth of the World Wide Web and the improved accessibility of information has exacerbated the problem of information overload in our daily lives \cite{ho2001towards}. In particular, this becomes an issue when trying to process and understand the barrage of news events \cite{shahaf2010connecting}. Furthermore, the rise of social media and its subsequent place as a staging ground for information operations \cite{weedon2017information, robbins2019weaponization} has given place to the creation of alternative narratives \cite{nied2017alternative} and the spread of fake news \cite{ciampaglia2018fighting}. 

Moreover, narratives play a key role as one of the basic building blocks of societal sensemaking \cite{wilson2018assembling} and political actors attempt to use strategic narratives to create a shared understanding of the world \cite{miskimmon2017forging}. Thus, given the importance of narratives in society, it is critical to understand how they emerge and evolve over time in this environment of information overload and fake news. 

Existing work has focused on algorithms to extract timelines \cite{shahaf2010connecting}, trees \cite{ansah2019graph}, or other variants of graphs to represent stories \cite{yang2009discovering}. However, they have mostly neglected the theoretical foundations of narratives on their models. The present work is an attempt to bridge this theoretical gap by creating a computational representation model for narratives grounded in formal narrative theory. In particular, the proposed model is based on a route map metaphor and it links elements from narrative theory and graph theory in an intuitive way.

\subsection{Narrative Theory}
The design of our narrative representation is informed by concepts from general narrative theory and strategic narrative theory. Both approaches contain elements that help us define and analyze our narrative representation.

General narrative theory focuses explicitly on understanding the general rules of narrative and its different arrangements that make it meaningful \cite{abbott2008cambridge, puckett2016narrative}. The key intuition in formal narrative theory is that there is a distinction between the story itself and its representation. Narrative theory tries to understand the relationships between stories and their many possible representations \cite{puckett2016narrative}.

In contrast, instead of focusing on the narrative itself, strategic narrative theory studies the construction and application of these narratives by political actors. In particular, strategic narrative theory focuses on studying how political actors communicate narratives in a strategic way to construct a shared meaning of events with the purpose of influencing the behavior of domestic and international actors \cite{miskimmon2014strategic,roselle2014strategic}. 

Another approach to narratives by Halverson et al. \cite{halverson2011master} defines narratives not just as one story, but rather as a system of stories. That is, narratives are a systematic collection of interrelated stories with coherent themes. We rely on this definition of narrative as a system of interrelated stories to build our narrative map extraction method. In particular, our method also ensures a certain degree of thematic coherence between the extracted storylines.

\subsection{Existing approaches to extract stories and narratives}
To inform the design of our extraction method, we refer to the existing story and narrative representations found in the literature. First, we discuss related works based on the level of detail (\textit{representation resolution}), for example, events or topics in the narrative. Afterward, we analyze related works based on the internal structure of the representation (\textit{representation complexity}), for example, a tree or a timeline.

\subsubsection{Representation Resolution}
Multiple resolution levels could be studied in the context of story and narrative visualization. In particular, most of the story visualization research has focused on either the event level \cite{shahaf2010connecting, liu2017growing} or the topic level \cite{nallapati2004event,kim2011topic,zhou2017survey}. Event-level approaches focus on concrete sequences of events. In contrast, topic-level approaches are more abstract. Thus, instead of focusing on the specific events that happen throughout the story, they focus on the overarching topics and how they relate to each other. 

In particular, previous studies have used graph representations based on event clustering to represent topics \cite{nallapati2004event}. Our method builds upon the idea of using clusters of events to represent topics, in particular, we use them to extract coherent storylines aligned with these clusters.

Some authors have also proposed more fine-grained resolution levels to represent stories, such as individual named entities \cite{faloutsos2004fast}, claims and attributions made in an article \cite{soni2014modeling}, or mixed resolution approaches that allow zooming between the different levels, such as the information cartography approach from Shahaf et al. \cite{shahaf2013information}. 
In contrast to these fine-grained approaches, we focus on the event level to build our narrative map representation. By choosing this resolution level, we are able to easily develop a direct mapping between narrative and computational elements.

\subsubsection{Representation Complexity}
There exist many approaches to visualize the evolution of news events. The simplest cases correspond to a linear representation of the events (i.e., simple sequences of events), such as timelines \cite{wang2015socially, yan2011evolutionary} and coherent story chains \cite{shahaf2010connecting, shahaf2012connecting}. All these representations consist of a single starting and ending event with a sequence of events joining them. 

Previous research on timeline and story extraction relies on optimizing different criteria to select events. Furthermore, these criteria could depend on the type of data used by the extraction method. For example, in the socially informed timelines approach by Wang et al. \cite{wang2015socially}, their optimization method seeks to balance topical cohesion between articles and comments in social media. In contrast, in the news timeline summarization method of Yan et al. \cite{yan2011evolutionary}, the optimization algorithm tries to balance relevance, coverage, coherence, and cross-date diversity. Another approach corresponds to the \textsf{Connect-the-Dots} algorithm by Shahaf and Guestrin \cite{shahaf2010connecting}, which creates story chains using a linear programming approach to maximize the coherence of the timeline. An improved version of their approach \cite{shahaf2012connecting} used integer programming to generate the story chain. Building upon the intuition of these previous works, our approach also relies on an optimization problem to select the events that appear in our narrative representation.  

A logical extension to the single timeline approach is to consider more than one timeline. These timelines can be shown in parallel in the story map. Such as in the work done by Xu et al. \cite{xu2013summarizing}, where they extract multiple storylines surrounding a news event. Due to the complexity of generating such maps by brute force, they develop a near-optimal algorithm based on locality-sensitive hashing to perform dimensionality reduction. They display all the storylines in the same story map as parallel timelines. In contrast to this approach, our method allows for the different storylines in the narrative to interact with one another, allowing them to merge or separate into new storylines. This allows a more appropriate representation of narratives as a system of interrelated stories, rather than disjointed timelines.

More advanced cases use directed acyclic graph structures, such as event evolution graphs \cite{yang2009discovering, zhou2017emmbtt}, and metro maps \cite{shahaf2012trains,shahaf2013information}. In these cases, the maps can have multiple starting and ending events. Furthermore, these structures allow for more advanced local substructures, such as convergence (merging stories) and divergence (separating stories). Scholars have studied different models and features to represent and extract the relationships between events in these graphs. For example, the event evolution graph of Yang et al. \cite{yang2009discovering} uses event timestamps, event content similarity, temporal proximity, and document distributional proximity to model relationships. In contrast, Zhou et al. \cite{zhou2017emmbtt} use a model based on term and event frequency, alongside event sequences, content similarity, and temporal distance costs. Another approach corresponds to optimizing the connectivity of the graph, subject to coherence and coverage constraints \cite{shahaf2012trains}. Using these We model our event relationships using a similarity-based approach in conjunction with clustering information to ensure topical coherence and a certain degree of coverage. 

Finally, we also have tree-based approaches to represent stories. For example, the tree-based representation of Ansah et al. \cite{ansah2019graph} leverages information propagation among users in a community, temporal proximity, and semantic context to construct coherent paths and build a timeline summary. In addition, the news content organization system of Liu et al. \cite{liu2017growing} build a forest representation using keyword extraction, event clustering, story clustering, coherence, and time-based penalties.

\section{An Algorithmic Approach to Represent and Extract Narratives}
\label{sec:representation}
We offer a graph-driven approach to represent and find narratives from large troves of unstructured data corresponding to real-world events. We start by showing our computational representation of narratives (RQ1). Next, we present our algorithmic approach to extract these narratives automatically from data (RQ2). 

\subsection{RQ1: Narrative Representation}
We offer a conceptual representation of narratives while synthesizing elements from the narrative theory literature and following the route map metaphor. Table \ref{tab:minimal-narrative-map} lays out the key elements in a narrative, their metaphoric equivalents, and their respective computational representations.

\begin{table}[tb]
\hyphenpenalty=10000\exhyphenpenalty=10000
\scriptsize
\renewcommand{\arraystretch}{1.3}
\sffamily
\centering
\begin{tabular}{@{}p{2cm}p{2cm}p{2cm}p{7cm}@{}}
\toprule
\multicolumn{3}{c}{\textbf{Elements}} & \multicolumn{1}{c}{\textbf{Description}} \\ \cmidrule(r){1-3}
\textbf{Narratological} & \textbf{Computational} & \textbf{Metaphorical} & \multicolumn{1}{c}{} \\ \midrule
Events & Nodes & Landmarks & Events are actions or happenings that are part of the narrative.  \\
Discourse Elements & Node Attributes & Landmark Attributes & Discourse Elements represent additional or complementary information about the events that are not an event of the story (e.g., the news source that published the event). \\
Starting Event & Source & Starting Place & The first event of a story. All stories must begin at one of these events and there must be at least one such event on the map.  \\
Ending Event & Sink & Ending Place & The last event of a story. All storylines terminate in an end event and there must be at least one such event on the map.\\
Story & Chain (Path) & Route & A story is a sequence of events and discourse elements. Under the definition of narrative as a system of stories, there can be multiple stories in the same narrative.\\
Coherence & Edge Weights & Evaluation Metric & Coherence represents how much sense it makes to join two events and serves as an evaluation metric for the route. \\
Constituent Events & Maximum Likelihood Chain & Main Route & Constituent events form the core events of the narrative and are part of the main storyline. \\
Supplementary Events & Maximum Antichain & Representative Landmarks & Supplementary events correspond to events that are not the core part of the narrative but that provide further information and are part of alternative storylines. \\
\midrule
Narrative & st-graph & Route Map & The narrative is a system of interrelated stories, we represent the narrative as an st-graph. \\ 
\midrule
\multicolumn{4}{c}{\textbf{Visual Representation}}\\
\multicolumn{4}{c}{\includegraphics[width=0.85\textwidth]{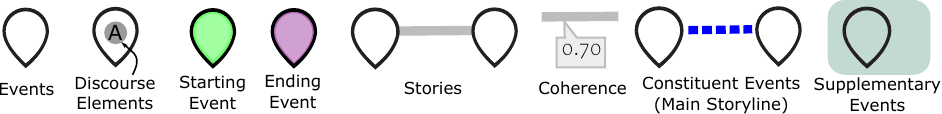}}\\
\bottomrule
\end{tabular}
\caption{Mapping of narrative theory elements and computational elements alongside their route map metaphor equivalent. The concepts shown here are used interchangeably throughout the paper to describe our algorithmic approach. Narratives have at least one storyline, and a storyline has at least two events. Thus, we have the hierarchy: event $<$ storyline $\leq$ narrative. }
\label{tab:minimal-narrative-map}
\end{table}

\subsubsection{What are the key elements in a narrative?}
Synthesizing across theories of narratives, we compile eight elements fundamental toward forming a narrative. Most of these elements are based on or related to the concept of events. For example, stories are sequences of events, coherence represents how much sense it makes to join events, and constituent events are a special kind of event that is fundamental to the progress of the narrative. All these elements rely on the basic concept of event for their definition, but they represent a specific narratological concept \cite{abbott2008cambridge}. The narrative itself is computationally represented as a graph, with each narrative element corresponding to a building block of this graph.

\noindent\paragraph{Story, Events, and Discourse Elements} According to Abbott's general narrative theory \cite{abbott2008cambridge}, narratives are composed of two primary elements: \emph{story} and \emph{discourse}. The story is a sequence of \emph{events} involving entities and follows a chronological order. Thus \emph{events} are one of the fundamental building blocks of a \emph{story}.
Meanwhile, the narrative \emph{discourse} is the story as narrated, that is, the rendering of the story in a particular narrative (in other words, a representation of the story).
While \emph{events} correspond to actions or happenings in the story, \emph{discourse} elements are part of the representation of the story. Discourse elements provide additional framing or complementary information about the events that influence the way the narrative is told but are not events themselves. Examples include public reactions to an event on social media or attributes of an event, such as the credibility of the source reporting the event or bias in reporting. 

Thus, in formalizing the computational equivalent of these elements, we represent events as nodes in a graph, discourse elements as node attributes (e.g., an attribute containing the event source), and the graph itself corresponds to the narrative map. We unpack the computational representation of a narrative by introducing two other elements---starting and ending events. 

\noindent\paragraph{Starting and Ending Events}
General narrative theory \cite{abbott2008cambridge} further states that a narratological story is defined by a single starting and ending event. Some authors argue that only one event is needed to have a story \cite{smith1980narrative, genette1983narrative, abbott2008cambridge}, but many concur that at least two connected events are essential \cite{worth2008storytelling, tversky2004narratives, rimmon2003narrative, barthes1975introduction}. We align with the popular view and require our narrative map to contain at least two events: \textit{starting event} and \textit{ending event}. 
We represent the starting event as a source in the graph---a node with no incoming edges. Likewise, we represent the ending event as a sink---a node with no outgoing edges. As per the route map metaphor, these two events correspond to the starting and ending places in a physical map, respectively. As in the physical map, the starting and ending events are selected by the user.

\noindent\paragraph{Story Coherence and System of Stories}
What makes a story coherent?  
A narrative story comprising of events and discourse elements needs to flow together, i.e., it needs to be coherent \cite{abbott2008cambridge}. For example, a narrative story that focuses on a certain topic should not drastically change to a different topic without appropriate events joining them. In other words, \emph{coherence} represents how much sense it makes to join two events. Thus, we can computationally represent story coherence through the weight of the edge joining the two events, where events are represented as nodes in a graph; higher edge weight implies higher coherence. Starting with this representation, next in RQ2, we offer computational techniques to automatically find the edge weight and determine whether to join the two events. Story coherence, thus, corresponds to the evaluation metric of our route map metaphor.

Moreover, prior theoretical work posits ``narrative as a system of stories'' \cite{halverson2011master}. Following this definition, our narrative map should not only contain a single story but should represent a narrative as a system of stories. These \emph{stories} must be interrelated and form coherent sequences of events. 

\noindent\paragraph{Constituent and Supplementary Events}
An important part of analyzing a narrative is distinguishing which events are the core and which ones simply provide complementary information to the narrative \cite{abbott2008cambridge}. 
According to theories of narrative, \textit{constituent events} represent the core. These are the main events that drive the story forward and cannot be omitted without fundamentally altering the story. In contrast, \textit{supplementary events} do not drive the main story forward and if removed, the main story would still remain intact. However, this does not imply that supplementary incidents are not necessary to understand the context and general impact of the narrative. Hence, we allow both constituent and supplementary events as key elements for our narrative representation.

In computationally formalizing these two narratological elements, we want the overall sequence of constituent events to make as much sense as possible. Thus, we represent constituent events as nodes along the most coherent path (i.e., the maximum likelihood path) in the narrative map that joins the starting event with the ending event. The most coherent path connecting the sequence of constituent events also provides us with the main storyline. In terms of the route map metaphor, the sequence of constituent events corresponds to the ``main route'' that has the best value for our evaluation metric (i.e., coherence) out of all the routes. For example, recall the blue route in Figure \ref{fig:metaphor}(b), which reflects the development of the outbreak from its first case as a mysterious virus to the US banning flights to Europe. Note in particular, that his route has a likelihood of $0.7\times 1.0 \times 0.5 \times 0.3 = 0.105$, the highest out of all routes in the map.

For supplementary events, we focus on finding a representative subset of these events. To do this, we focus on all potential alternative storylines that join the starting event with the ending event. However, enumerating all potential alternative storylines is computationally impractical, except for the smallest of the narrative maps. Instead, we extract meaningful representative events from each alternative storyline in the narrative map. In terms of the route map metaphor, each alternative route can be associated with a representative landmark (i.e., a supplementary event). While these routes do not have the highest likelihood, they can still be viable alternatives and can offer different perspectives. For instance, recall the example from our route map metaphor (Figure \ref{fig:metaphor}(a)) where we elaborated alternative routes, such as a longer scenic route through the Chill Pond versus the preferred main route through the South Market.

\subsubsection{How can we define a structured way to navigate across the narrative elements?}
Now that we have formalized the basic building blocks of the narrative representation, we define the underlying computational structure that relates them. 
Recall that a narrative map is computationally represented as a graph. However, to ensure that our representation aligns with the theoretical structure of narratives, we must specify additional constraints to this graph structure. According to narrative theory, narratives can arrange events in many ways \cite{genette1983narrative}, but the underlying stories are made up of \textit{sequences} of events that follow \textit{chronological order} \cite{abbott2008cambridge}. 
To capture these properties, we use a directed acyclic graph (DAG) because the event nodes must be joined in sequences (directed) and they must follow chronological order (acyclic). Moreover, this structure also fits the requirements of the route map metaphor, which does not contain cycles and provides a sequential order for the landmarks in a route. 

However, there is one caveat of having a general DAG with no additional constraints. A DAG can have multiple sources (i.e., nodes with no incoming edges) and sinks (i.e., nodes with no outgoing edges). Yet, our route map metaphor only contains one source and one sink (the starting and ending place). Thus, we need to find a graph structure that respects these constraints. Specifically, we impose two additional restrictions on our DAG representation: the graph must have exactly one source node and one sink node. This is a special kind of graph called an \textit{st-graph} \cite{husfeldt1995fully}---a directed acyclic graph with a single source and a single target (sink). See for example the graph of Figure \ref{fig:metaphor}(b), where the direction of the edges is implicitly given by the chronological order of the events. 
This representation allows us to encode a single system of stories connecting the ending and starting events (i.e., a single narrative map).
By using an \textit{st-graph} rather than a general directed acyclic graph, the representation allows us to understand how the events are connected without considering unnecessary information about other starting and ending events. Furthermore, we incorporate the coherence measure as edge weights in our graph. Finally, this structure allows us to navigate throughout the elements of the narrative map, offering us a chronological view of all the events and storylines contained in it.

\subsubsection{Formalizing the narrative map definition}
Given our previous discussion, we propose the following formal definition for a narrative map:
\begin{definition}
\textbf{Narrative map}: We formally define a narrative map as a weighted directed acyclic graph $G = (D, E)$ with a single source $s \in D$ (the starting event) and a single sink $t \in D$ (the ending event), such that weights represent the probability of going from one node to another based on coherence and the paths from $s$ to $t$ represent the different storylines in the narrative.
\end{definition}

Figure \ref{fig:metaphor}(b) from the motivational example section shows an example of a weighted \textit{st-graph} for a sample narrative, where the direction of the edges is implicitly given by the chronological order of the events. It should be noted that our representation is general enough, such that, if we modify some of the constraints, we can obtain other representations. For instance, a simple timeline (one that only allows sequential patterns, such as a single story chain \cite{shahaf2010connecting}) or an event tree (one that allows at most one parent for each node) can be constructed from our generic representation. As a case in point, to convert our narrative map in Figure \ref{fig:metaphor}(b) to a narrative tree, we could require that nodes have at most one parent, relax the restriction of only one ending event, and then drop the less important edges. 
More complex representations, such as parallel timelines \cite{xu2013summarizing} or story forests \cite{liu2017growing}, can also be obtained by combining the simpler representations.

\subsection{RQ2: Narrative Map Extraction}
Here we introduce our narrative map extraction method. We first define the narrative map problem formally before presenting the narrative extraction process. Figure \ref{fig:process} gives an overview.

\subsubsection{Formal Problem Definition}
Based on our previous formalization of a narrative map as a directed acyclic graph with a single source (the starting event) and a single sink (the ending event), we provide a formal definition of the narrative map problem.

\begin{definition}
\textbf{Narrative map problem}: Given a set of time-stamped documents $D$ representing events for a certain issue of interest (e.g., news articles for the Coronavirus outbreak), a starting event $s \in D$ and an ending event $t \in D$, our objective is to find a graph in the form of a narrative map $G = (D, E)$, where source $s$ and sink $t$ are connected (by potentially many paths). 
\end{definition}

Since the narrative map problem has multiple potential solutions, we need to define the criteria to find an appropriate solution---a good narrative map.

\subsubsection{What makes a good narrative map?}
To answer, we rely on the notion of narrative coherence \cite{shahaf2010connecting, yan2011evolutionary, shahaf2012trains}---how much sense it makes to join two events---and coverage \cite{yan2011evolutionary, shahaf2012trains,shahaf2013information}---how well the map covers the different topics in the narrative. 

Inspired by previous works \cite{shahaf2013information,wang2015socially,liu2017growing,zhou2017emmbtt}, we base our measure of coherence on the following intuition: coherence should be high if the events are similar and if they share common topics. That is, in a narrative map when a user follows a chain of events along edges joining them, they should get a clear understanding of the underlying theme in the storyline.
Therefore, to determine whether it makes sense to join two events with an edge on the narrative map graph, we need to compute both event similarity \cite{zhou2017emmbtt} and topic similarity \cite{shahaf2013information,wang2015socially,liu2017growing}.

Based on previous works \cite{shahaf2012trains,shahaf2013metro}, we base our measure of coverage on the following intuition: coverage should be high if the map covers as many topics of the narrative as possible. Prior work has relied on simple word frequency-based models and pre-defined keyword lists to compute coverage and give priority to important words \cite{shahaf2012trains}.
However, a pre-defined keyword-based approach makes assumptions about how events will look a priori, potentially increasing the risk of selection bias \cite{mitra2015credbank}. Thus, to avoid this issue we use a more abstract approach by defining coverage with respect to clusters of events (representing topics) that are automatically extracted from data.

In general, we want narrative maps to be as coherent as possible while also covering the topics in an appropriate way. Thus, we regard a narrative map as good if it is maximally coherent for a given coverage requirement. This leads naturally to an optimization approach, which we cover in more detail in the next subsection. 

\subsubsection{Narrative Extraction Process}
Following our previous discussion on what makes a good narrative map, we seek to maximize the coherence of the narrative map, subject to coverage and structural constraints. First, we give a high-level overview of our optimization method and the extraction process. Then, we provide additional details for each phase of our algorithmic approach starting from data extraction and ending with the final visualization. 

\paragraph{Extraction Process Overview}
We develop an optimization-based approach informed by prior scholarly work \cite{shahaf2010connecting}. In particular, we employ a linear programming technique which automatically generates a narrative map by selecting a subset of events and connections between them, such that it maximizes coherence subject to two constraints: 1) coverage constraints to ensure that our solution has at least a certain level of coverage and 2) structural constraints to ensure that our solution has an st-graph structure.

Based on this optimization approach, our extraction process consists of 6 phases. The input corresponds to a data table with headlines, publication dates and URLs. The output corresponds to a visualization representing the narrative through the route map metaphor. In phase 1 we extract the data and perform basic preprocessing. In phase 2, we compute coherence information for each pair of events. Coherence values are then used as input to the linear program. In phase 3, we solve the linear program to obtain the final st-graph representation. Note that during this phase we compute coverage and ensure that our solution meets the required threshold. In phase 4, we extract the main route from the st-graph representation by finding the route with the highest coherence. In phase 5, we extract the representative landmarks from the st-graph representation using graph antichains---sets of pairwise incomparable nodes that can be used to represent the storylines. Finally, in phase 6, we generate the final narrative map and its visualization based on the route map metaphor.

\begin{figure}[!htbp]
    \centering
    \includegraphics[width=0.93\textwidth]{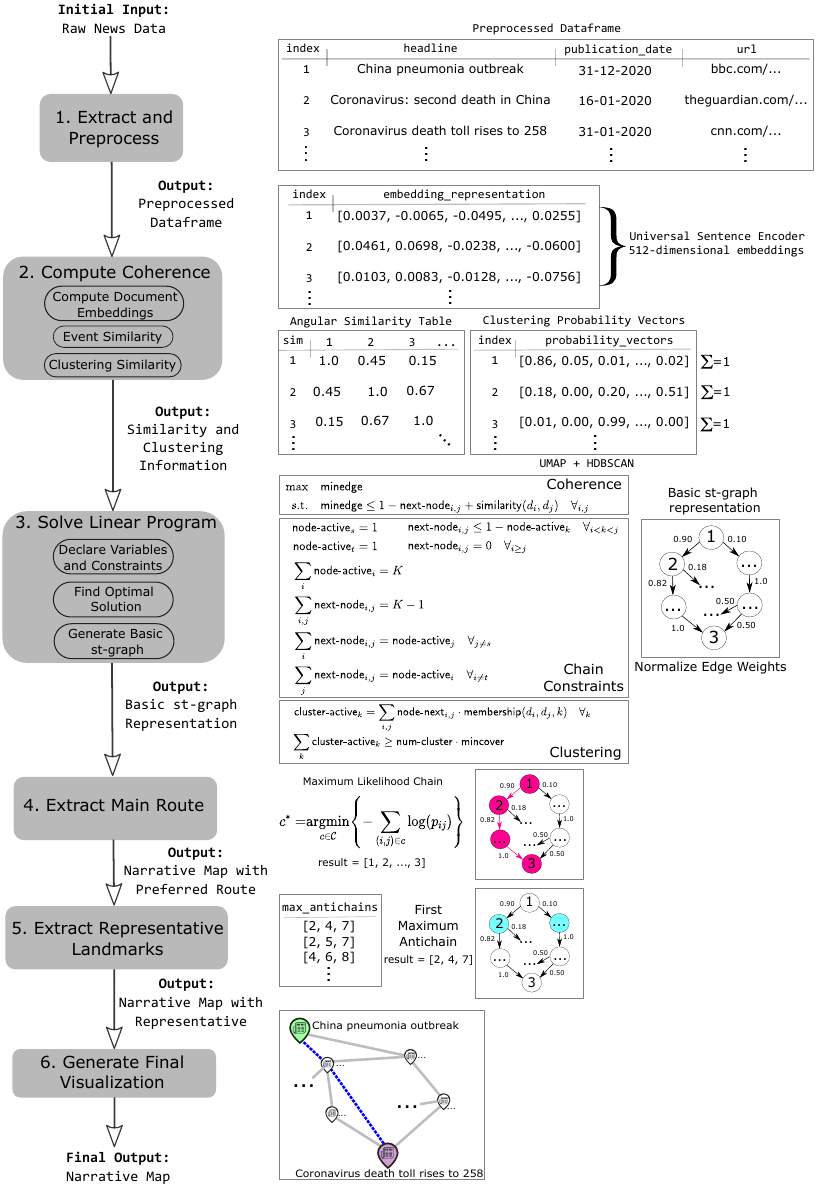}
    \caption{Figure illustrating the steps in our narrative extraction algorithm that generates a narrative map from data (RQ2). Each phase is annotated with their respective outputs.}
    \label{fig:process}
\end{figure}

\paragraph{Phase 1: Extracting Data and Preprocessing}
Our data set of choice is a collection of news articles. News articles capture real-world information and are structured to provide the key event details at the very beginning in the news headline \cite{norambuenaevaluating}. Thus, we retain headlines of the news articles as our unit of analysis. Each headline corresponds to an event in the narrative. Next, we perform basic data preprocessing steps, such as fixing character encoding issues. 

\paragraph{Phase 2: Computing Coherence}
Coherence comprises of two components---event similarity and topic similarity.
First, to compute event similarity, we represent the events (corresponding news headlines in our case) using neural embeddings of the headlines. In particular, for each event, we used the pre-trained models from the Universal Sentence Encoder v4.0 from TensorFlow \cite{cer2018universal} to extract the document embeddings from their headlines. This representation is appropriate for our purpose because it has been trained on news data and is designed for short text \cite{cer2018universal}. After mapping news headlines on the document embedding space, we find semantically similar headlines by comparing the similarities between pairs of embeddings. Instead of traditional raw cosine similarity, we use angular distance-based similarity measure which distinguishes near parallel vectors better and is well suited when used with the Universal Sentence Encoder \cite{cer2018universal}. Angular similarity ranges between 0 and 1; values close to 1 denote similar events. 

Next, to compute the topic similarity component, we obtain potential topics through clustering, where each cluster corresponds to an underlying topic. In particular, we cluster the documents using HDBSCAN \cite{mcinnes2017hdbscan} with UMAP \cite{mcinnes2018umap}. HDBSCAN is an advanced density-based technique, but it scales poorly with the dimensionality (number of features) of the data. Thus, to improve the speed and performance of HDBSCAN, we use UMAP on the document embeddings for dimensionality reduction as a preprocessing step. In addition, an important advantage of HDBSCAN is that it does not require specifying the number of clusters (or topics) a priori, allowing us to automatically surface the most important topics directly from data. The step results in cluster probability vectors, containing the probability of whether a certain document belongs to each cluster.

Next, we need to compare the topic similarity of two events using their cluster probability distributions. We compare the cluster distribution vectors using the Jensen-Shannon similarity (JS-score)---a preferred method to compare probability distributions \cite{hadjila2019flexible}. If the events have perfectly equal topic or cluster distribution vectors, we obtain a JS similarity score of 1 (e.g., both events have $[0.6, 0.3, 0.1]$ as its cluster distribution vector). If the events focus on completely different storylines or topics, the JS similarity score will be 0 (e.g. $[0.4, 0.6, 0]$ vs. $[0.0, 0.0, 1.0]$).

How can we combine the two components---event similarity and topic similarity---to obtain the final coherence score? A simple approach would be to compute the arithmetic mean. However, this approach is not strict enough for our purposes. We want a scoring mechanism that has strong penalties if any of the components are low and forces a strict balance them.
Thus, we rely on the geometric mean (square root of their product), which imposes stricter penalties when a score is low. In particular, by formulating it through a geometric mean rather than a simple arithmetic mean, we ensure that the coherence will be zero if and only if any of the components are zero. 

\paragraph{Phase 3: Solving the Linear Program}
Using our coherence information from the previous step, we can now implement the linear program that maximizes coherence. The solution that we generate from this linear program is an \textit{st-graph} that represents interrelated stories, thus fitting with the definition of narrative as a system of stories from Halverson et al. \cite{halverson2011master}. The mathematical formulation of the linear program is shown in the corresponding phase of Figure \ref{fig:process}. Here we explain each component of this formulation in an intuitive way. The linear program seeks to optimize the strength of the weakest link in the narrative map in terms of coherence.  Intuitively, this is based on the idea that a story is as coherent as its weakest link \cite{shahaf2010connecting}. Furthermore, recall that narratives have a single starting event, a single ending event, and are formed by sequences of events ordered chronologically, represented as an \textit{st-graph}. Thus, our optimization problem needs to include a series of structural constraints to ensure that our solution is an \textit{st-graph}. Namely, it needs to set up a single source (starting event) and a single sink (ending event) and then force the events to be joined chronologically, These are all characteristics of a directed acyclic graph (DAG for short). 

Finally, the last constraint---coverage constraints---ensures that a certain percent of the extracted topic clusters are covered. We found empirically that a minimum coverage of 80\% worked well for our data in preliminary qualitative evaluations. Using this parametrization, we found that in general, the extracted storylines made sense and the covered topics were relevant. Furthermore, we note that coverage, unlike coherence, can not be pre-computed because it can change with each iteration of the linear program optimization process (different potential solutions of the linear program can yield different coverage values).

\begin{figure}[tb]
    \centering
    \includegraphics[width=1.0\textwidth, keepaspectratio]{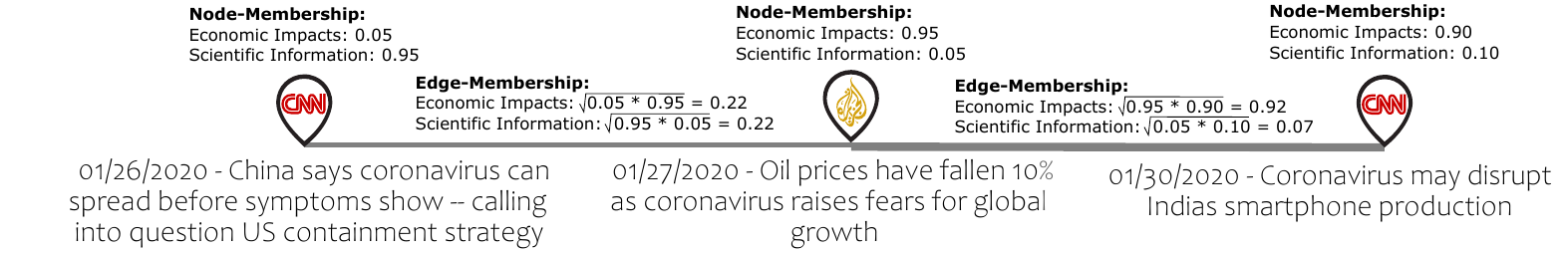}
    \caption{Overview of the edge membership computation used in the coverage constraint. The example shows three events from the Coronavirus narrative and two clusters. The first edge does not contribute much to the coverage of either cluster, because it is joining nodes from different clusters. In contrast, the second edge joins nodes from the same cluster, and thus it strongly contributes to the economic impacts cluster.}
    \label{fig:membership-example}
\end{figure}

To compute coverage, we rely on the following intuition: a topic is covered if it appears on coherent sequences of events. Following this intuition, nodes on their own are not enough to consider a topic covered, because they might be isolated events sprinkled throughout the map. Instead, we rely on edges to determine if a topic is properly covered. Consider for example the storyline in Figure \ref{fig:membership-example} comprised of three events. This storyline focuses mostly on the economic impact of the virus. However, the first event \textit{[China says Coronavirus can spread before symptoms show (...)]} is mostly unrelated to economic impacts, instead, it talks about scientific information. Since this event is isolated in a storyline about a completely different topic (economic impacts) and does not connect with other events talking about its own topic (scientific information), it would not meaningfully contribute to the scientific information topic coverage.

Computationally, we define the coverage of the topic cluster $k$ as the sum of all edge weights that belong to that cluster weighted by edge membership to cluster $k$ (see for example Figure \ref{fig:membership-example}). The membership of an edge to a cluster is defined as the geometric mean of the membership of each endpoint. We use the geometric average because its properties work intuitively well with the concept of membership. Similar to what we did with the coherence computation, we want a scoring mechanism that has strong penalties if any of the endpoints does not belong to the cluster. 

For example, if both endpoints have a non-zero probability of being in that cluster, their edge gets a positive membership. Otherwise, if any of them has a zero probability, the edge membership is also zero. Consider the storyline in Figure \ref{fig:membership-example} based on the Coronavirus narrative, with only two clusters we can clearly see the impact of joining two nodes from distinct clusters to the values of edge membership. Furthermore, while the first edge does not contribute much [0.22] to the coverage of either of the clusters, the second edge strongly contributes [0.92] to the economic impacts cluster. We normalize the outgoing edge weights for each node to ensure that they sum up to 1. This turns our coherence weights into probabilities, which helps to interpret them.

\paragraph{Phase 4: Extracting the Main Storyline}
Thus far, our algorithmic approach has constructed the basic narrative map structure. Now we use this structure to extract the main route or storyline (i.e., the constituent events). To do this, we find the most coherent storyline from the starting event to the ending event in the narrative map. In other words, we find the path with the highest probability from the source to the sink, or equivalently, the maximum likelihood path. In general, we can find the probability of each path by multiplying the edge weights. However, instead of maximizing the multiplication of probabilities, in practice, it is more convenient to minimize the sum of negative \textsf{log}-probabilities \cite{forsyth2015improving}. By turning this into an additive minimization problem, we can use shortest path algorithms on our graph with \textsf{log}-probability weights to find the maximum likelihood path.

In practice, there is usually only one such route, but it is theoretically possible to have more. If there are multiple routes, this would imply that the algorithm has produced two potential sets of constituent events to describe the narrative and is incapable of choosing the best one.

\paragraph{Phase 5: Finding Representative Landmarks with Antichains}
Now that we have found the main route, we next find the representative landmarks in the narrative map. Before delving into the method, we introduce some important concepts from graph theory that informed our technique.

\begin{figure}[tb]
    \centering
    \includegraphics[width=0.2\textwidth, keepaspectratio]{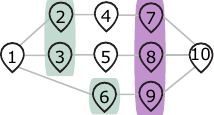}
    \caption{Antichain examples. The nodes highlighted in green form the first maximum antichain of the graph with 3 elements. The nodes highlighted in purple also form a maximum antichain, but it is not the earliest antichain. The nodes 2 and 3 on their own also form an antichain, but it is not a maximum antichain. Note how each element of the antichain is associated with a chain (left, center, and right)}
    \label{fig:antichain}
\end{figure}

To extract the representative landmarks, we leverage the concept of maximum antichains in directed acyclic graphs. Formally, antichains are sets of pairwise incomparable nodes in the DAG (i.e., nodes that are not connected by a path) \cite{pijls2013another}. For example nodes 2 and 3 form an antichain in Figure \ref{fig:antichain}. In particular, an antichain is called \textit{maximum} antichain if no other antichain has more nodes. For example, nodes 2, 3, and 6 form one possible maximum antichain, no other antichain that we select can have more nodes.

Previous works have shown that maximum antichains formalize the concepts of abstract graph views, slices \cite{gansner2005topological}, or abridgments (i.e., a small part of the graph that is of interest) \cite{eades2004navigating}. Furthermore, researchers have used them to define a partitioning of the nodes of the graph \cite{abello2006ask}. In particular, according to Dilworth's theorem \cite{dilworth2009decomposition}, the size of the maximum antichain---called the width of the DAG---is equal to the minimum number of chains (paths) into which we can partition the DAG. Intuitively, we can partition the graph into chains (i.e., storylines) and interpret each node of the maximum antichain as a representative element for one of the chains. This intuitive interpretation is backed by Dilworth's formal theorem. Figure \ref{fig:antichain} demonstrates this with an example. Take the chains of Figure \ref{fig:antichain}, $A: 1\rightarrow2\rightarrow4\rightarrow7\rightarrow10$, $B: 1\rightarrow3\rightarrow5\rightarrow8\rightarrow9$, and $C: 1\rightarrow6\rightarrow9\rightarrow10$, we can associate one element from an antichain to each of these chains, such as $(A,2), (B,3), (C,6)$.

To find the maximum antichains, we use the antichain extraction method implemented in the \textsf{NetworkX} library \cite{hagberg2008exploring} in Python. This gives us a list with all the antichains (i.e., list of node indices). We filter the list, ensuring that it only contains maximum antichains. Note that there are potentially multiple maximum length antichains. In this case, our tiebreaker strategy is to select the earliest antichain according to their timestamps, as this provides us with representative landmarks close to the starting points of each storyline. For example, consider the two maximum antichains in Figure \ref{fig:antichain}. In that case, we would choose the green antichain for our representative landmarks. 

\subsubsection{Phase 6: Generating the Final Visualization}
Finally, we turn our computational representation into a visual representation, which is used for the case study and user evaluation. In particular, we use GraphViz \cite{ellson2004graphviz} and the DOT language \cite{gansner1993technique} to generate the layout of our visualization in an SVG file. Then, we embed the SVG file into an HTML page and add basic interactive elements using CSS. 

Our visual representation follows the route map metaphor that we described in the introduction. The metaphor elements in Table \ref{tab:minimal-narrative-map} guide our design and we try to faithfully follow our original concept in the illustrations of Figure \ref{fig:metaphor}. The visualization shows the date and the headline accompanied by a landmark icon with the news source logo. We note that our visualization omits the edge weight information. We removed it because preliminary evaluations proved that adding explicit weights confused the users. Instead, we make edges wider to represent higher weights.

\begin{figure}[!htbp]
    \includegraphics[width=0.972\textwidth, keepaspectratio]{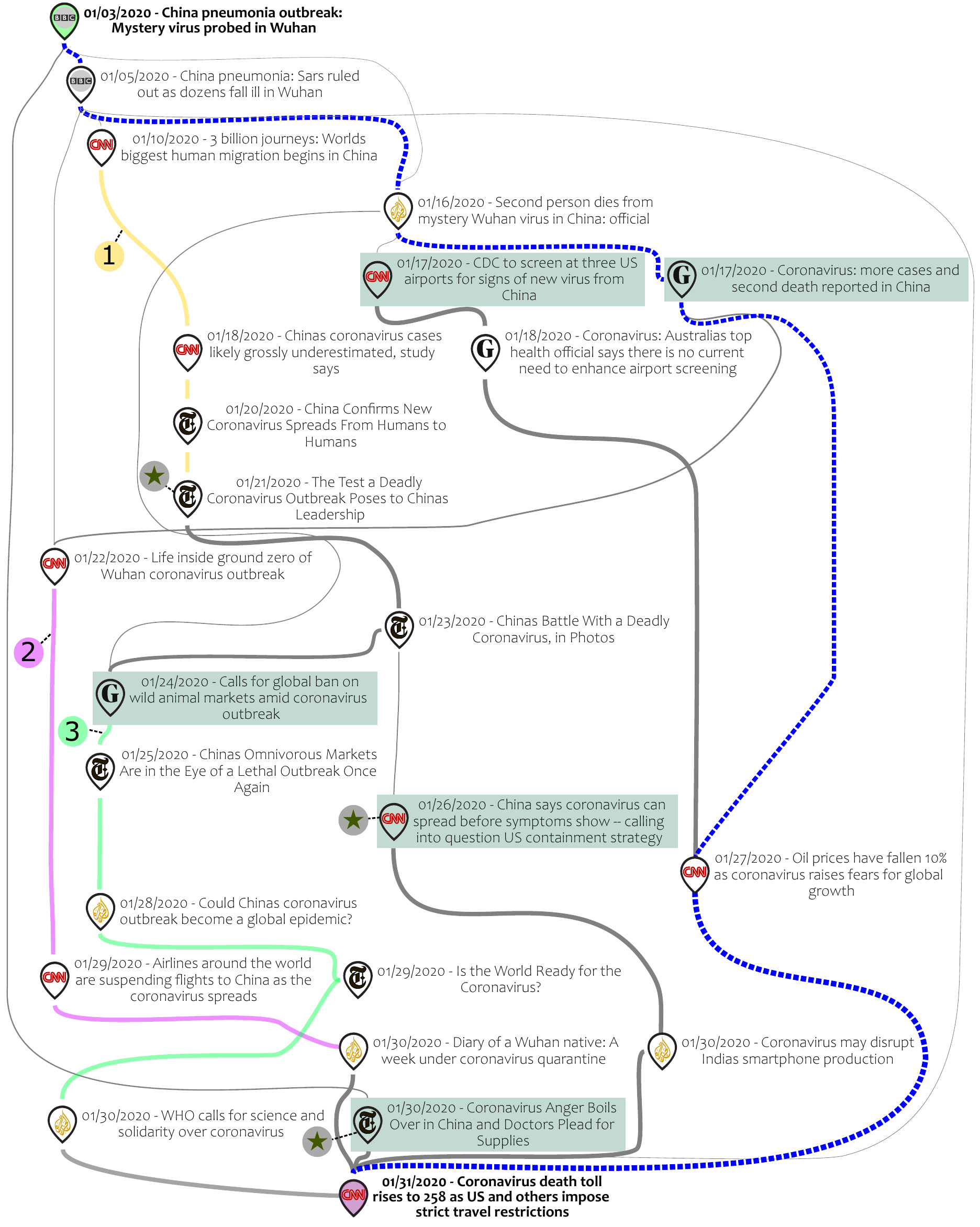}
    \caption{Narrative Map for the Coronavirus Outbreak during January. Headlines in bold denote the starting event \protect\resizebox{0.30cm}{!}{\protect\includegraphics{figures/start.pdf}} and ending event \protect\resizebox{0.30cm}{!}{\protect\includegraphics{figures/end.pdf}}. The blue dashed line corresponds to the main storyline \protect\resizebox{0.35cm}{!}{\protect\includegraphics{figures/main_route.pdf}} (the maximum likelihood path). The width of each edge depends on its weight (coherence). Events highlighted in green are representative landmarks \protect\resizebox{0.35cm}{!}{\protect\includegraphics{figures/representative_landmarks.pdf}} (members of the first maximum antichain). The markers \protect\resizebox{0.35cm}{!}{\protect\includegraphics{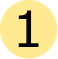}}, \protect\resizebox{0.35cm}{!}{\protect\includegraphics{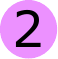}}, and \protect\resizebox{0.35cm}{!}{\protect\includegraphics{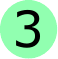}} highlight some storylines that we reference throughout our discussion, we display them in distinctive colors. The \protect\resizebox{0.35cm}{!}{\protect\includegraphics{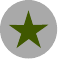}} markers show events that share the theme of questioning or criticizing government responses.}
    \label{fig:january}
\end{figure}

\section{Case Study: Coronavirus Narrative}
In this section, we showcase our narrative extraction algorithm in action. We present a case study to demonstrate how we can leverage the narrative route map representation and our computational extraction method to understand the narrative as it evolves and changes. First, we present the data used to generate the narrative map. Then, we show our results and analysis.

\subsection{Overview of the Case Study}
We focus on the Coronavirus outbreak due to its high impact and how clearly defined the major events are (e.g., first case, lockdown in Wuhan, the closing of borders, stay at home orders). Note that it is not a requirement to have such clearly defined major events to create a narrative map, but it facilitates the analysis and thus it makes for an informative example. 

We obtained data related to the Coronavirus outbreak from the COVID-19 archive\footnote{https://www.covid19-archive.com/}. This archive maintains a list of news articles related to the outbreak and comprised 1576 articles from 373 unique sources spanning three months---Jan'20 to Mar'20. For the purposes of this case study, we focused on the first month of the outbreak and retained articles from December 31st to January 31st, resulting in a data set of 607 news articles.

To reduce the search space of the optimization algorithm, we selected articles from the top 5 news domains in terms of publications (BBC, Al Jazeera, The Guardian, CNN, and New York Times), resulting in 102 news articles from the first month of the outbreak. Focusing on the major sources allows us to obtain an overview of the most prominent storylines and the major themes in the narrative while ignoring less important stories from more obscure sources. 
For the starting and ending events, we simply took the chronological first and last elements in the corpus. Out of the 102 articles, our narrative extraction algorithm returned an optimal solution with 22 events, in addition to the fixed starting and ending events. The rest were not selected because they lowered the overall coherence when added to any of the extracted storylines. Figure \ref{fig:january} shows our results. 

\subsection{Analysis: Narratives underlying the Coronavirus Outbreak}

\begin{figure}[tb]
    \centering
    \includegraphics[width=1.0\textwidth, keepaspectratio]{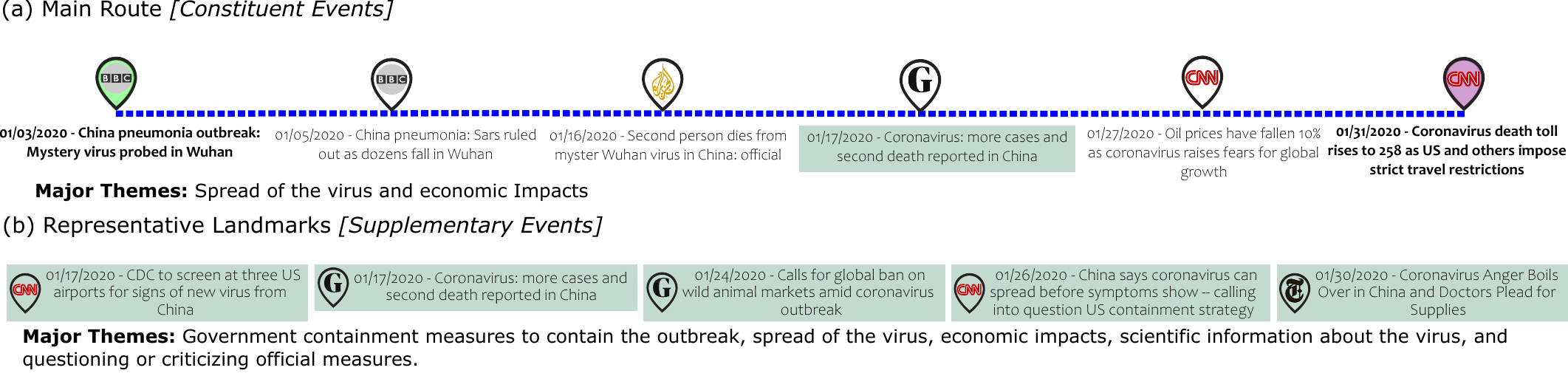}
    \caption{Analysis summary of the main route and representative landmarks for the January narrative map. We extract the major themes out of these elements.}
\label{tab:summary-jan}
\end{figure}

What are the narratives present in the first month of the Coronavirus outbreak? Figure \ref{fig:january} shows the full narrative map generated by our algorithm. Figure \ref{tab:summary-jan} shows a summary of the major themes of the narrative according to the main route \resizebox{0.35cm}{!}{\includegraphics{figures/main_route.pdf}} and the representative landmarks \resizebox{0.35cm}{!}{\includegraphics{figures/representative_landmarks.pdf}}. 

We begin by analyzing the \textbf{main route} in Figure \ref{tab:summary-jan}(a), which starts by referring to the new ``mysterious pneumonia virus'' in Wuhan. At this point in the outbreak, information about the virus is scarce, but at the very least SARS was ruled out. Then, less than a week later, there are already two deaths from this mystery virus. The next day news outlets report the second death and the virus has now been called ``Coronavirus''. After this point, the story changes its focus and talks about the drop in oil prices as the Coronavirus raises fears for global growth. Finally, this leads to the final event with a death toll of 258 and the US and other countries imposing restrictions. Thus, the major themes of the main route are the general spread of the virus and its economic impacts.

\begin{figure}[tb]
    \centering
    \includegraphics[width=1.0\textwidth, keepaspectratio]{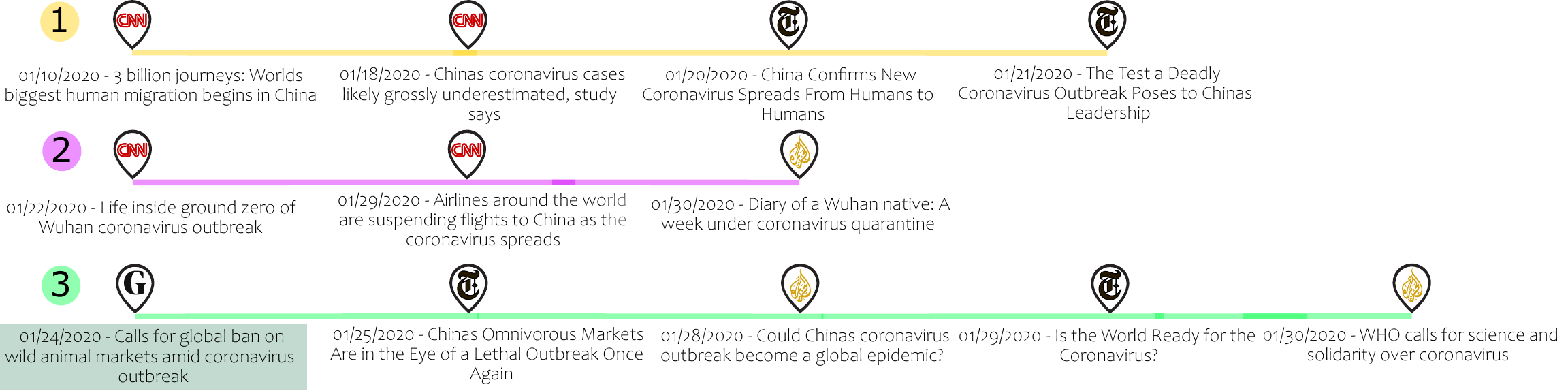}
    \caption{Examples of substories from the Narrative Map for the Coronavirus Outbreak during January. For brevity the start event \protect\resizebox{0.30cm}{!}{\protect\includegraphics{figures/start.pdf}} and ending event \protect\resizebox{0.30cm}{!}{\protect\includegraphics{figures/end.pdf}} are not shown in detail. Events highlighted in green are representative landmarks \protect\resizebox{0.35cm}{!}{\protect\includegraphics{figures/representative_landmarks.pdf}}. We use markers \protect\resizebox{0.35cm}{!}{\protect\includegraphics{figures/1.pdf}}, \protect\resizebox{0.35cm}{!}{\protect\includegraphics{figures/2.pdf}}, and \protect\resizebox{0.35cm}{!}{\protect\includegraphics{figures/3.pdf}} to refer to the storylines as in Figure \ref{fig:january}.}
    \label{fig:substories}
\end{figure}

The \textbf{representative landmarks} in Figure \ref{tab:summary-jan}(b) reveal a thematic overlap with the main route as well as surface new elements, where common topics overlapping with the main route include the ``spread of the virus'' and its ``economic impact.'' New themes in the representative landmarks refer to ``scientific information about the virus'' \textit{[China says Coronavirus can spread before symptoms show (...)]}, ``government measures to contain the outbreak'' \textit{[CDC to screen at three US airports (...)]}, and ``criticisms and questioning of government responses'' \textit{[Coronavirus anger boils over in China (...)]}.

Next, we turn our attention to some of the \textbf{stories} on the map. For instance, see \resizebox{0.35cm}{!}{\includegraphics{figures/1.pdf}} in Figures \ref{fig:january} and \ref{fig:substories}. Here we see references to mass migration as people try to escape from the virus. Then we find out that the number of cases might be grossly underestimated and that humans can transmit the virus to other humans. Thus, this storyline implies that the virus is likely to spread from human mass migration.

Another storyline deals with the lockdown in Wuhan and the surrounding provinces (see  \resizebox{0.35cm}{!}{\includegraphics{figures/2.pdf}} in Figures \ref{fig:january} and \ref{fig:substories}). At this point, we see multiple articles about how life goes on during the quarantine. This storyline focuses on a more human aspect of the outbreak, seeing how it affects the daily lives of people. This is a theme that we did not identify directly by looking at the main route and the representative landmarks. This implies that some themes are abstracted away when we look at the narrative exclusively from the main route and representative landmarks perspective. 

Amidst the battle with the virus, we see a route dealing with wild animal markets (see  \resizebox{0.35cm}{!}{\includegraphics{figures/3.pdf}} in Figures \ref{fig:january} and \ref{fig:substories}), one of the potential causes of the outbreak according to the information available at that time. Then, this turns into a discussion about whether the virus could become a global pandemic and whether the world is ready to handle it. This is followed by the World Health Organization (WHO) asking for science and solidarity over the virus. 

There are also events dealing with anger over how the Chinese government is handling the situation, talks about the challenges the outbreak could pose to the Chinese government, and questioning the United States containment strategy. These events are part of different storylines (see the \resizebox{0.35cm}{!}{\includegraphics{figures/star.pdf}} markers in Figure \ref{fig:january}), yet they share a common theme in terms of questioning governments and how they are handling the outbreak.

\section{User Evaluation}
Finally, to provide additional support for the effectiveness of our narrative map representation and answer RQ3, we perform a user evaluation. Here, we present the guidelines driving our evaluation followed by the results of our user study.

\subsection{Evaluation Guidelines}
The narrative map representation should be an effective method to communicate complex narratives to individuals. Hence, to measure the effectiveness of the representation, the underlying metaphor, and the overall algorithmically generated narrative-map visual, we evaluate three dimensions:

\begin{enumerate}
	\item \textbf{Representation}: How well does the map represent the narrative?
	\item \textbf{Metaphor}: How well does the route map metaphor work for narratives?
	\item \textbf{Visualization}: How well does the visualization work in general?
\end{enumerate}

\begin{table}[tb]
\hyphenpenalty=10000\exhyphenpenalty=10000
\scriptsize
\sffamily
\begin{tabular}{@{}lp{3.5cm}p{5.0cm}p{2cm}@{}}
\toprule
\textbf{Dimension}      & \textbf{Element}               & \textbf{Questionnaire Item}                                                                   & \textbf{Source}                \\ \midrule
\textbf{Representation} & Coherence                      & The map presents a coherent overview of the narrative.                                        & Shahaf (2010)    \\
                        & Relevance                      & The map presents relevant information about the narrative.                                    & Shahaf (2010)    \\
                        & Redundancy                     & The map has too much redundant information.                                                   & Shahaf (2010)    \\
                        & Effectiveness                  & I feel more familiar with the narrative after using this map.                                 & Shahaf (2010)    \\
                        & Completeness                   & The amount of information on the map is appropriate to represent the narrative.               & Burkhard (2005)        \\
                        & Size                           & The size of the map is appropriate to represent the narrative.                                & Burkhard (2005)        \\ \midrule
\textbf{Metaphor}       & Nature                         & The metaphor is natural (i.e., it is similar to the real-world).                              & Garc\'ia (2015) \\
                        & Understanding                  & The metaphor is understandable (i.e., easy to understand and extract information).            & Garc\'ia (2015) \\
                        & Interpretability               & It is easy to draw conclusions and get new insights from the metaphor.                        & Garc\'ia (2015) \\
                        & Landmarks                      & Presenting narrative events as landmarks makes sense.                                         & Burkhard (2005)        \\
                        & Main Route                     & The main storyline serves well as an overview of the most important events in the narrative. & New                            \\
                        & Representative Landmarks            & The representative landmarks serve as an overview of all the stories in the narrative.        & New                            \\ \midrule
\textbf{Visualization}  & Visual Impact                  & The visualization is eye-pleasing.                                                            & Shamim  (2015)    \\
                        & Overall Performance            & The visualization is easy to understand.                                                      & Shamim (2015)    \\
                        &                                & The visualization is user-friendly.                                                           & Shamim (2015)    \\
                        & Overall Design Style           & The visualization is informative.                                                             & Shamim (2015)    \\
                        &                                & The visualization is intuitive.                                                               & Shamim (2015)    \\
                        & Information Quality            & The visualization is useful.                                                                  & Shamim (2015)    \\
                        &                                & The comprehensiveness of the data is good.                                                    & Shamim (2015)    \\
                        & Visual Representation Model    & The visualization allows us to easily compare storylines.                                     & Shamim (2015)    \\
                        &                                & The representation style of the data is good.                                                 & Shamim (2015)    \\
                        & Information Presentation Model & Previous knowledge is required to understand the visualization.                               & Shamim (2015)    \\ \bottomrule
\end{tabular}
\caption{Table containing the evaluation dimensions and their elements, as well as the questionnaire items and their source (if applicable). Some questions taken from a source have been adapted to our metaphor. The questionnaire has a total of 22 elements. Subjects were asked to evaluate each statement according to a 5-point Likert scale, ranging from ``Strongly Disagree'' to ``Strongly Agree.''}
\label{tab:questions}
\end{table}

For each of these dimensions, we adapt evaluation questionnaires from a range of prior scholarly work. Table \ref{tab:questions} shows an overview of the questionnaire. For example, to evaluate along the representation dimension, we adopt Shahaf et al's \cite{shahaf2010connecting} evaluation scheme from their Connect-the-Dots algorithm---an algorithm that heavily informed our extraction approach. We also take elements from Burkhard and Meier's work \cite{burkhard2005tube}, where they present a framework for evaluating knowledge visualization tools that are based on visual metaphors. To evaluate the effectiveness of our metaphor, we adopted elements from Garc\'ia et al.'s \cite{garcia2015visualisation} work, where they focused on visualization metaphors to process complex information and interpret data. Finally, we also take elements from the visualization evaluation  questionnaire of Shamim et al. \cite{shamim2015evaluation}---a survey questionnaire designed to assess the usability of visualization systems pertaining to news articles. To measure the internal consistency of our final 22-item evaluation questionnaire, we carried out a reliability analysis. Cronbach's alpha with $\alpha = 0.9560$ demonstrates the high internal consistency of our evaluation questionnaire.

\subsection{Evaluation Procedure}
Can users with no training in narrative or news analysis use and interpret our algorithmically generated narrative map? To answer, we asked Amazon Mechanical Turk workers (MTurkers) to do short tasks. Recent work by Cheng et al. \cite{cheng2019explaining} has shown the promise in employing MTurkers in the context of explaining complex AI algorithms, where they leverage the results from an MTurk evaluation with non-expert users to guide the design of their systems. Moreover, prior scholarly works have used MTurk in intelligence analysis tasks and have obtained high-quality results,  demonstrating the capabilities of MTurk \cite{li2019dropping}. For example, MTurkers have performed at par with intelligence analysts in connecting entities in a visualization map \cite{li2018crowdsourcing} and in identifying relationships in large sets of data\cite{neigel2018role}. Therefore, even if MTurkers are not a replacement for expert feedback, they could still act as good proxies and provide valuable insights towards how our narrative maps proposal representation, metaphor, and visualization work in the context of sensemaking and intelligence analysis tasks. Appendix A contains snapshots detailing the MTurk job. 

Prior research \cite{paolacci2010running, hauser2018common} indicates that it is possible to obtain good quality data from MTurk. Specifically, as long as we follow certain guidelines, the data quality is similar to traditional sources (e.g., student recruitment pools). Furthermore, research using MTurk faces many ethical challenges \cite{williamson2016ethics, gleibs2017all}, in particular with regards to payment. These challenges have given rise to various initiatives to tackle them, such as Dynamo \cite{salehi2015we}. We have followed the general Dynamo guidelines for academic requesters in MTurk. We have ensured that our MTurk workers receive appropriate compensation, above the federal minimum wage according to the estimated response time. Furthermore, we provided workers with an explanation in case they were rejected due to failing attention check questions. 

The task required that MTurkers spent time to interact with the visualization, understand the structured data it surfaced, infer one of the major themes of the main route and two themes of the whole narrative map (different from the main route one), followed by responding to our 22-item evaluation questionnaire (shown in Table \ref{tab:questions}). 
Subjects were required to answer how much they agreed with the statements shown in Table \ref{tab:questions} on a 5-point Likert scale with the following options: Strongly Disagree, Disagree, Undecided, Agree, and Strongly Agree. 
In addition to the 22 questions, we added one attention check question. Three workers responded incorrectly to the attention check question. We discarded their answers and requested additional workers to keep our total to 20 valid responses for each narrative map task evaluation. The entire duration of the task along with reading instructions was approximately 30 minutes. Workers were compensated for \$4 for each task, adhering to federal minimum wage standards.

To warrant high-quality responses, we required MTurkers to be Master Turkers, have at least a 95\% approval rating, and to have completed at least 50 tasks on MTurk. 
To ensure that the workers correctly understood the task, we first trained them by providing an explanation of the basic elements of the narrative route map metaphor. We also showed them a simple example narrative map with just three nodes, including a short discussion of the common themes that could be extracted from the example map. Prior work in crowdsourcing suggests that training workers by showing them examples results in high-quality responses \cite{lasecki2014glance, mitra2015comparing}.

We performed our MTurk evaluation with three different narrative maps, each requiring 20 responses. All the maps were focused on the Coronavirus outbreak but spread across three different months to facilitate the evaluation: January, February, and March 2020.

\begin{figure}[tb]
    \centering
    \includegraphics[width=0.9\textwidth]{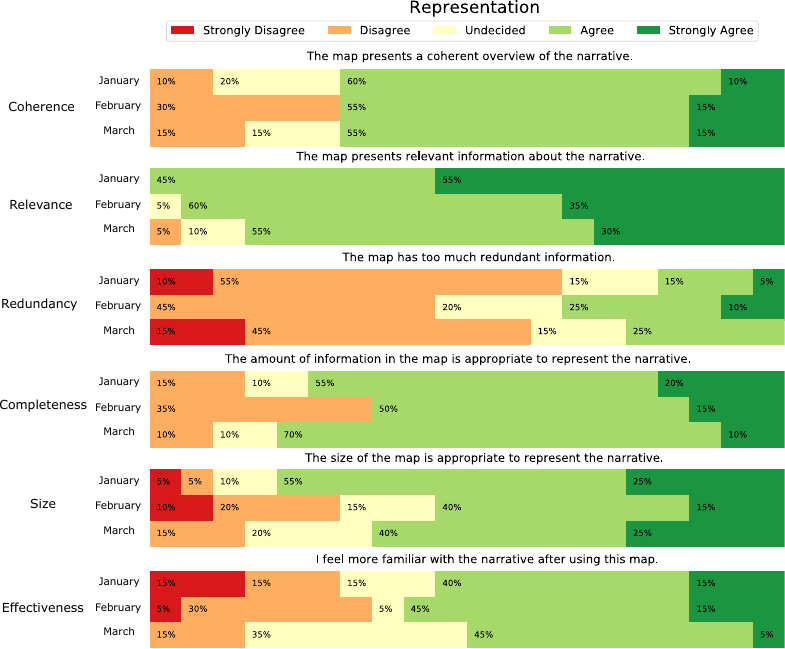}
    \caption{User evaluation results for the representation dimension. The bars show the percentage of users per response for each question.}
    \label{fig:rep_eval}
\end{figure}

\begin{figure}[tb]
    \centering
    \includegraphics[width=0.9\textwidth]{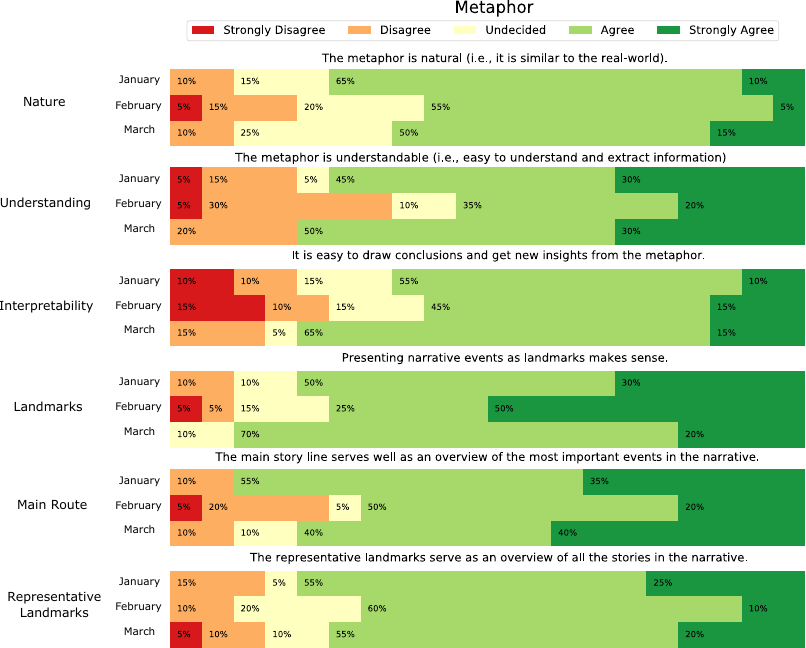}
    \caption{User evaluation results for the metaphor dimension. The bars show the percentage of users per response for each question.}
    \label{fig:rep_metaphor}
\end{figure}

\begin{figure}[tb]
    \centering
    \includegraphics[width=0.9\textwidth]{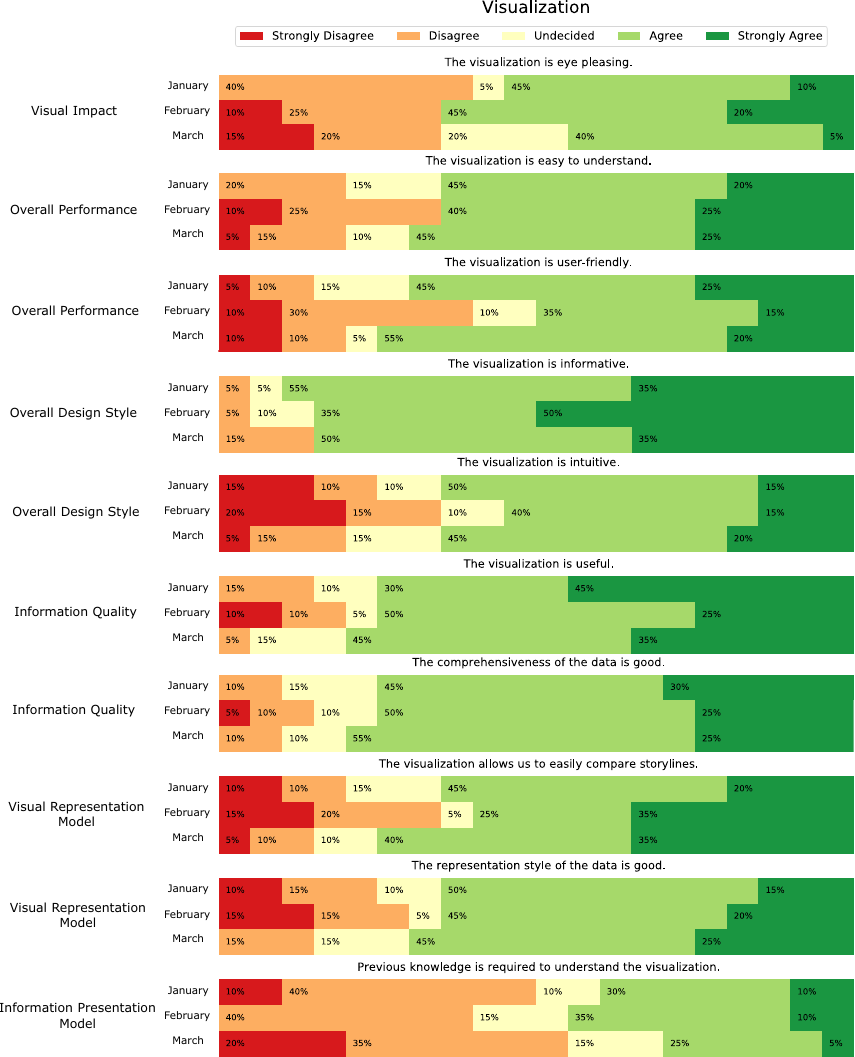}
    \caption{User evaluation results for the visualization dimension. The bars show the percentage of users per response for each question.}
    \label{fig:rep_vis}
\end{figure}

\subsection{Evaluation Results}
Here, we present the results of our user study across each dimension (representation, metaphor, and visualization). Overall, all dimensions were evaluated well by the MTurkers as shown in Figures \ref{fig:rep_eval}, \ref{fig:rep_metaphor}, and \ref{fig:rep_vis}. These results are mostly consistent throughout the three months. For brevity, we focus on a few interesting results from our analysis. 

\subsubsection{Representation}
First, we analyze the results of the representation dimension. Overall, the maps are considered as a \textit{coherent} overview of the narrative for each month. Next, we note that there is a slight drop in \textit{relevance} during March, which might be because the March map has 40 events and is significantly larger than the other maps. Thus, there might be too many irrelevant events compared to the other maps. In contrast, we see that most users did not consider our representation \textit{redundant}. This result is encouraging considering that one of the potential issues of using a similarity-based approach was that our representation could be redundant since it would naturally keep trying to connect highly similar items. In terms of \textit{size}, we see that March had better results than February, despite being the larger map. This difference in results could be due to layout differences rather than the number of nodes itself. Moreover, we see that there is a trade-off between an appropriate map size and \textit{completeness} of the data. In this context, and considering our results, our narrative extraction method balances both of these requirements adequately. Finally, we see that at least half of the users felt more familiar with the topic after using the visualization, thus showing the \textit{effectiveness} of our representation. While we expected a more positive response to this question, it should be noted that due to how pervasive Coronavirus news was at the time we did our user study, it is possible that Turkers were already very well acquainted with the information at hand. With a more obscure topic, the visualization could have had a greater impact on the users by providing new knowledge.

\subsubsection{Metaphor}
Second, we analyze the results of the metaphor dimension. In general, the results for this dimension were positive. Users agree that our route map metaphor as well as the more specific metaphors (\textit{landmarks}, the \textit{main route}, and \textit{representative landmarks}) are indeed an intuitive and \textit{natural} way to represent a narrative. Furthermore, the users agree that it is easy to draw conclusions from the metaphor, thus it is highly \textit{interpretable}. Moreover, most users found the metaphor to be \textit{understandable} for January and March. However, fewer users agreed to this for the February map. We believe that this is due to the complex layout of the February map compared to the other two. 

\subsubsection{Visualization}
Finally, we turn to our analysis of the visualization dimension. Overall, the evaluation results for this dimension were positive as well. In terms of \textit{visual impact}, most users found the visualization to be eye-pleasing, except for March. This difference could be due to the size of the map. In terms of its \textit{overall performance}, the visualization had good results in both ease of understanding and user-friendliness. Regarding the \textit{overall design style}, users mostly agreed that the current design is informative and intuitive. Likewise, in terms of \textit{information quality}, the visualization had good results regarding its usefulness and the comprehensiveness of the data. The \textit{visual representation model} was also evaluated well by the users in the context of comparing storylines. Finally, with respect to the \textit{information presentation model}, around half of the Turkers considered that the narrative representation did not require previous knowledge. The rest of the users believed that previous knowledge would be required to understand the visualization. This split in the user evaluation indicates that additional measures might be needed to facilitate using the visualization without prior knowledge. However, it might also reflect the fact that most people likely had previous knowledge about Coronavirus due to the constant barrage of news surrounding this topic. We leave additional evaluations, with a more obscure topic, for future work.

\section{Discussion}
Our work provides a computational representation of narratives grounded in formal narrative theory. We also offer an algorithmic approach to extract these representations from data using a linear programming technique. We summarize our approach and outline the key implications. 

\subsection{RQ1: Narrative Representation}
Unlike previous attempts at extracting storylines or narratives from data \cite{shahaf2010connecting,shahaf2013information, yan2011evolutionary,wang2015socially,zhou2017emmbtt}, our work is based on a comprehensive survey of narrative theory. Hence our narrative representation provides a general framework to represent narratives and can serve as a starting point for understanding narratives through a computational lens.

Our representational framework emulates the real-world metaphor of a route map, making it more usable and understandable while displaying abstract unstructured knowledge (also demonstrated by our user evaluation results). Moreover, the computational representation of our framework---graph-based---can be effortlessly extended to encompass other narrative phenomena. 
For example, we can augment our data set to contain social media information such as comments, tweets, or posts related to our news events. By feeding this additional data to our narrative building model, we can incorporate the reactions of people towards certain articles. In turn, this provides us with insights into the issue of narrative reception \cite{roselle2014strategic} (i.e.,  how the public perceives narratives) and how it relates to the projected narrative by the media. 

Additionally, we could also add new discourse elements in the form of contextual information through text summaries and data plots. For example, in the case of the Coronavirus narrative, our narrative map could be augmented by including a text summary of the outbreak during that month and traditional data plots to show the growth of the virus as the narrative advances (see Figure \ref{fig:future}).

\subsection{RQ2: Algorithmically Extracting Narratives}
The optimization-driven approach for generating narrative maps uses an objective function that seeks to maximize the coherence of the narrative based on similarity and clustering information. The resulting solution of this optimization problem selects real-world events that are displayed in the narrative map along with their connections. Our user evaluation study shows that our approach constructs a primarily coherent and relevant narrative from hundreds of news articles.

Moreover, we include a coverage constraint to force the narrative map to cover many topics. This constraint could be modified and expanded to consider different kinds of coverages, similar to the idea of user-personalized coverage from previous scholarly work \cite{shahaf2013metro}. For example, by adding a term to consider the political bias of the news sources, we could require a certain level of coverage for each side of the political spectrum. Thus, ensuring that all sides of the story are covered. 

To help users interpret the storylines, major themes, and the connection between events, we provide them with the main route and representative landmarks of our map. These elements offer a way to characterize narratives by helping identify the storylines and major themes of the narrative. Our user study confirmed that both the main route and the representative landmarks served well in providing an overview and highlighting key elements of the narrative. While the \emph{main route} supplies with a \emph{longitudinal view} of the narrative, \emph{representative landmarks} created by the maximum antichain offer a \emph{transverse view}. In essence, the representative landmarks and the main route provide us with an understanding of both the breadth and the depth of the narrative, respectively.

\subsection{RQ3: Evaluation}
The evaluation proves that the Narrative Map representation, alongside the extraction method, is an effective mechanism to help users understand complex news narratives. The route map metaphor was considered natural by the users, providing a coherent overview of the narrative. Moreover, the main route and the representative landmarks are able to convey the key elements of the narrative.

Although the user evaluation demonstrated the effectiveness of our method, there is room for improvement. We focus in particular on the possibilities for enhancing the visualization method. As noted in the user evaluation, it is necessary to refine its visual appeal. Moreover, reducing the need for previous knowledge in the visualization is an important consideration. This could be achieved by including auxiliary discourse elements to provide additional context (e.g., news summaries and complementary data visualization).

Further improvements could be made by adding more interactivity to the visualization. For example, a simple tweak would be to highlight all edges of a node when an edge is selected. More complex upgrades could allow users to move the nodes along with its edges or allow them to hide or merge certain nodes while preserving structure. For instance, an approach similar to the recursive summarization technique for discussion forums \cite{zhang2017wikum} could be adopted for narratives, enabling a user to explore different narrative storylines based on their interests, with varying levels of details. Investigating these extensions could be fruitful avenues for future research. Nevertheless, in its current state, the visualization correctly represents the narrative and is able to provide users with new insights, as shown by our user evaluation study with MTurkers.

\subsection{Implementation Parameters and Constraints}
Our linear programming problem has a total number of variables in the order of $O(|D|^2\cdot k)$, where $|D|$ is the number of documents and $k$ is the number of clusters. In practical terms, the number of clusters is small enough to be considered constant. Thus, our main computational bottleneck lies in generating the pairwise combinations of nodes required by the optimization problem. Future work could deal with improvements to the linear programming approach or even applying a completely different optimization technique, similar to how the Connect-the-Dots method \cite{shahaf2010connecting} evolved from its basic linear programming formulation into a divide-and-conquer approximation approach \cite{shahaf2015information} 

We also note that there are two parameters of interest in our linear program proposal: the expected length of the main route ($K$, based on the Connect-the-Dots formulation \cite{shahaf2015information}), and the minimum average coverage of clusters ($\mathsf{mincover}$). Tuning these parameters is required to obtain a good narrative map representation. Moreover, since our method depends on document embeddings and clustering, the choice of the vector embedding space representation, the clustering algorithm, and all the corresponding parameters for clustering can also influence the final representation. Thus, special care is required to choose an appropriate vector representation for the events, as this can influence both the similarity and the clustering results. 

\subsection{Limitations}
One limitation of our study is that we constructed the event representation based on the headlines of the articles. Although headlines generally contain key event information, it would be useful to consider either the full text of the article or an appropriate summary. In particular, considering the headline and the lead of the article might be enough to capture the main event descriptors \cite{norambuenaevaluating}.

In terms of the user evaluation of our narrative maps, we only focused on the Coronavirus narrative with only 20 users for each map. In order to properly generalize our results, future work should focus on evaluating different narrative maps with more users and with different topics, ranging from well-known issues (e.g., Coronavirus) to more obscure ones. Nevertheless, our user evaluation provides an important baseline for future works involving this narrative representation. 

\subsection{Implications and Future Work}
\subsubsection{A General Framework to Extract and Represent Narratives}
Our method is grounded in the scholarly definitions of narrative theory. We start with abstract theoretical definitions and move toward concrete representations of the essential elements in a narrative. 
Hence, our representation should be able to capture any kind of narrative as defined in the narrative theory literature. 
As long as data is in the form of \textsf{(event, timestamp)} tuples, we can use our algorithm to construct a map representation. Moreover, we can generalize this approach to any kind of data provided it has a temporal ordering and that we can define a measure (i.e., the evaluation metric of our metaphor) that allows us to compute coherence through similarity and clustering. At a technical level, the generalizability of our approach stems from the abstract mathematical formulation used to extract the map, which could, in turn, be used for other types of data, such as image data (e.g., creating a narrative map out of pictures from an embedding representation and timestamps).

\subsubsection{Design Implications}
Grounded in theoretical constructs of narratology, our approach can automatically extract structured information from multiple data points and represent it as a narrative map---a visual metaphor similar to physical maps. We believe that this capability offered by our approach can be used in the design of a wide range of systems. For example, imagine a news reading tool that accompanies articles with a narrative map visualization to help readers understand the big picture of the narrative, or consider a social media monitoring system that shows how the online narrative surrounding a certain issue evolves over time. Indeed, literary theorists, while outlining the importance of narratives noted \cite{abbott2008cambridge}: ``\emph{we do not have any mental record of who we are until narrative is present as a kind of armature, giving shape to that record.}''

Furthermore, narrative maps could be used in the context of computational journalism, which focuses on providing tools for various computational needs in the journalistic context \cite{cohen2011computational}, such as topic detection, visualization, and sensemaking. 
Moreover, the theoretical basis of our representation, along with the general approach of extracting narratives from data, ensures that a designer can easily adapt our proposal to a different domain. For example, our model could be adopted by psychologists to help analyze personal narratives \cite{kohler2000analysis} or by narratologists to study literary works. 

\begin{figure}
    \centering
    \includegraphics[width=0.80\textwidth]{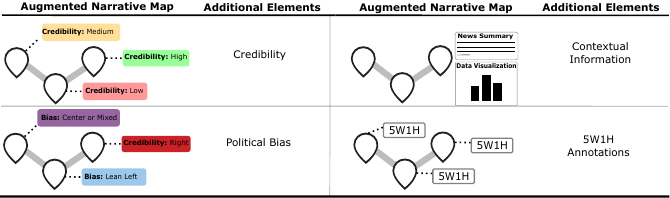}
    \caption{Potential extensions to narrative maps including different elements. The top-left map considers adding credibility annotations to each of the events, representing how credible or questionable the sources and their claims are. The bottom-left map considers adding political bias annotations, thus helping users understand how bias affects the storylines of the map. The top-right map considers additional information in the form of a news summary and data visualization. The bottom-right map considers adding the answers to the 5W1H questions (who, what, where, when, why, and how), which are the main descriptors of an event.}
    \label{fig:future}
\end{figure}

\subsubsection{Extending the Narrative Map Representation}
Using the foundations and methods presented in this paper, future work can create an enhanced representation, for example, by considering the bias and credibility of the sources of each event. See for instance how we added this information to the landmarks of the narrative maps in Figure \ref{fig:future}. With these enhanced maps, intelligence analysts would have an additional tool at their disposal to help detect potential online misinformation campaigns by analyzing emergent narratives. The rapidly shifting media ecology has made strategic narratives more difficult to control \cite{miskimmon2014strategic}, thus analysts would benefit from being able to detect counter-narratives as they emerge. Additionally, fact-checkers could use this enhanced representation to obtain an overview of how facts, both credible and questionable, spread throughout the narrative. This could help in their fact assessment process. Moreover, they could augment our representation by manually modifying it to add credibility annotations.

Furthermore, narrative maps can also be used as a summarization tool to identify high-level patterns in narratives. For example, for the Coronavirus narrative shown in our case study, we identified ``economic impacts'' and the ``spread of the virus'' as major themes. In conjunction with more advanced graph-theoretical elements, it would also be possible to identify more complex patterns in the narrative. For instance, we could use network motifs---frequent patterns in the structure of a graph \cite{milo2002network}---to encode interesting properties and relationships of its elements \cite{benson2016higher}.

\section{Conclusions}
This paper presents an algorithmic approach to represent and extract narratives using \textit{narrative maps}. Based on a comprehensive survey of formal narrative theory, we outline a computational representation of narratives comprising eight elements. We used an intuitive route map metaphor to connect the elements, their metaphoric renditions, and their computational counterparts. Building upon existing approaches of story extraction, we designed a novel optimization-based approach to extract the underlying graph representing a narrative map. In particular, we seek to maximize the coherence of the map, subject to structural and coverage constraints. Using this graph, we then extract the main storylines and the representative landmarks of the narrative map. These help us characterize the major themes in a narrative.

We presented a detailed case study following the Coronavirus outbreak during its first month. We explored the major events and identified the major themes surrounding the outbreak. This extended example helped us demonstrate how an analyst could leverage our representation to understand the evolution of a narrative. Next, we presented a user evaluation of our method considering three aspects: the narrative representation itself, the effectiveness of the metaphor, and the usability of the visualization. While there are still some aspects that could be improved, our evaluation results demonstrate that our narrative map proposal fulfills its intended purpose.

This work opens up multiple avenues for future research, such as adapting the current approach to another type of data (e.g., images) or to a different field (e.g., helping psychologists analyze personal narratives) or extending the narrative map representation by adding contextual information. In particular, incorporating measures of source bias and credibility into our representation could pave the way towards modeling how misinformation propagates in news narratives.

\begin{acks}
We would like to thank the social computing lab at Virginia Tech and University of Washington for their valuable comments and feedback on early drafts of the paper. This work was partially funded by NSF grants CNS-1915755 and DMS-1830501 and by ANID/Doctorado Becas Chile/2019 - 72200105.
\end{acks}

\bibliographystyle{ACM-Reference-Format}
\bibliography{acmart}


\begin{thebibliography}{74}


\ifx \showCODEN    \undefined \def \showCODEN     #1{\unskip}     \fi
\ifx \showDOI      \undefined \def \showDOI       #1{#1}\fi
\ifx \showISBNx    \undefined \def \showISBNx     #1{\unskip}     \fi
\ifx \showISBNxiii \undefined \def \showISBNxiii  #1{\unskip}     \fi
\ifx \showISSN     \undefined \def \showISSN      #1{\unskip}     \fi
\ifx \showLCCN     \undefined \def \showLCCN      #1{\unskip}     \fi
\ifx \shownote     \undefined \def \shownote      #1{#1}          \fi
\ifx \showarticletitle \undefined \def \showarticletitle #1{#1}   \fi
\ifx \showURL      \undefined \def \showURL       {\relax}        \fi
\providecommand\bibfield[2]{#2}
\providecommand\bibinfo[2]{#2}
\providecommand\natexlab[1]{#1}
\providecommand\showeprint[2][]{arXiv:#2}

\bibitem[\protect\citeauthoryear{Abbott}{Abbott}{2008}]%
        {abbott2008cambridge}
\bibfield{author}{\bibinfo{person}{H~Porter Abbott}.}
  \bibinfo{year}{2008}\natexlab{}.
\newblock \bibinfo{booktitle}{\emph{The Cambridge introduction to narrative}}.
\newblock \bibinfo{publisher}{Cambridge University Press},
  \bibinfo{address}{One Liberty Plaza, New York, NY, USA}.
\newblock


\bibitem[\protect\citeauthoryear{Abello, Van~Ham, and Krishnan}{Abello
  et~al\mbox{.}}{2006}]%
        {abello2006ask}
\bibfield{author}{\bibinfo{person}{James Abello}, \bibinfo{person}{Frank
  Van~Ham}, {and} \bibinfo{person}{Neeraj Krishnan}.}
  \bibinfo{year}{2006}\natexlab{}.
\newblock \showarticletitle{Ask-graphview: A large scale graph visualization
  system}.
\newblock \bibinfo{journal}{\emph{IEEE transactions on visualization and
  computer graphics}} \bibinfo{volume}{12}, \bibinfo{number}{5}
  (\bibinfo{year}{2006}), \bibinfo{pages}{669--676}.
\newblock


\bibitem[\protect\citeauthoryear{Ansah, Liu, Kang, Kwashie, Li, and Li}{Ansah
  et~al\mbox{.}}{2019}]%
        {ansah2019graph}
\bibfield{author}{\bibinfo{person}{Jeffery Ansah}, \bibinfo{person}{Lin Liu},
  \bibinfo{person}{Wei Kang}, \bibinfo{person}{Selasie Kwashie},
  \bibinfo{person}{Jixue Li}, {and} \bibinfo{person}{Jiuyong Li}.}
  \bibinfo{year}{2019}\natexlab{}.
\newblock \showarticletitle{A Graph is Worth a Thousand Words: Telling Event
  Stories Using Timeline Summarization Graphs}. In
  \bibinfo{booktitle}{\emph{The World Wide Web Conference}}
  \emph{(\bibinfo{series}{WWW ’19})}. \bibinfo{publisher}{ACM},
  \bibinfo{address}{New York, NY, USA}, \bibinfo{pages}{2565–2571}.
\newblock
\showISBNx{9781450366748}


\bibitem[\protect\citeauthoryear{Baber, Andrews, Duffy, and McMaster}{Baber
  et~al\mbox{.}}{2011}]%
        {baber2011sensemaking}
\bibfield{author}{\bibinfo{person}{Chris Baber}, \bibinfo{person}{Dan Andrews},
  \bibinfo{person}{Tom Duffy}, {and} \bibinfo{person}{Richard McMaster}.}
  \bibinfo{year}{2011}\natexlab{}.
\newblock \showarticletitle{Sensemaking as narrative: Visualization for
  collaboration}.
\newblock \bibinfo{journal}{\emph{VAW2011, University London College}}
  \bibinfo{volume}{VAW2011} (\bibinfo{year}{2011}), \bibinfo{pages}{7--8}.
\newblock


\bibitem[\protect\citeauthoryear{Barthes and Duisit}{Barthes and
  Duisit}{1975}]%
        {barthes1975introduction}
\bibfield{author}{\bibinfo{person}{Roland Barthes} {and}
  \bibinfo{person}{Lionel Duisit}.} \bibinfo{year}{1975}\natexlab{}.
\newblock \showarticletitle{An Introduction to the Structural Analysis of
  Narrative}.
\newblock \bibinfo{journal}{\emph{New Literary History}} \bibinfo{volume}{6},
  \bibinfo{number}{2} (\bibinfo{year}{1975}), \bibinfo{pages}{237--272}.
\newblock
\showISSN{00286087, 1080661X}


\bibitem[\protect\citeauthoryear{Benson, Gleich, and Leskovec}{Benson
  et~al\mbox{.}}{2016}]%
        {benson2016higher}
\bibfield{author}{\bibinfo{person}{Austin~R Benson}, \bibinfo{person}{David~F
  Gleich}, {and} \bibinfo{person}{Jure Leskovec}.}
  \bibinfo{year}{2016}\natexlab{}.
\newblock \showarticletitle{Higher-order organization of complex networks}.
\newblock \bibinfo{journal}{\emph{Science}} \bibinfo{volume}{353},
  \bibinfo{number}{6295} (\bibinfo{year}{2016}), \bibinfo{pages}{163--166}.
\newblock


\bibitem[\protect\citeauthoryear{Burke}{Burke}{1969}]%
        {burke1969grammar}
\bibfield{author}{\bibinfo{person}{Kenneth Burke}.}
  \bibinfo{year}{1969}\natexlab{}.
\newblock \bibinfo{booktitle}{\emph{A grammar of motives}}.
  Vol.~\bibinfo{volume}{177}.
\newblock \bibinfo{publisher}{Univ of California Press}, \bibinfo{address}{155
  Grand Ave. Suite 400. Oakland, CA}.
\newblock


\bibitem[\protect\citeauthoryear{Burkhard and Meier}{Burkhard and
  Meier}{2005}]%
        {burkhard2005tube}
\bibfield{author}{\bibinfo{person}{Remo~Aslak Burkhard} {and}
  \bibinfo{person}{Michael Meier}.} \bibinfo{year}{2005}\natexlab{}.
\newblock \showarticletitle{Tube Map Visualization: Evaluation of a Novel
  Knowledge Visualization Application for the Transfer of Knowledge in
  Long-Term Projects.}
\newblock \bibinfo{journal}{\emph{J. UCS}} \bibinfo{volume}{11},
  \bibinfo{number}{4} (\bibinfo{year}{2005}), \bibinfo{pages}{473--494}.
\newblock


\bibitem[\protect\citeauthoryear{Cer, Yang, Kong, Hua, Limtiaco, John,
  Constant, Guajardo{-}Cespedes, Yuan, Tar, Sung, Strope, and Kurzweil}{Cer
  et~al\mbox{.}}{2018}]%
        {cer2018universal}
\bibfield{author}{\bibinfo{person}{Daniel Cer}, \bibinfo{person}{Yinfei Yang},
  \bibinfo{person}{Sheng{-}yi Kong}, \bibinfo{person}{Nan Hua},
  \bibinfo{person}{Nicole Limtiaco}, \bibinfo{person}{Rhomni~St. John},
  \bibinfo{person}{Noah Constant}, \bibinfo{person}{Mario Guajardo{-}Cespedes},
  \bibinfo{person}{Steve Yuan}, \bibinfo{person}{Chris Tar},
  \bibinfo{person}{Yun{-}Hsuan Sung}, \bibinfo{person}{Brian Strope}, {and}
  \bibinfo{person}{Ray Kurzweil}.} \bibinfo{year}{2018}\natexlab{}.
\newblock \showarticletitle{Universal Sentence Encoder}.
\newblock \bibinfo{journal}{\emph{CoRR}}  \bibinfo{volume}{abs/1803.11175}
  (\bibinfo{year}{2018}), \bibinfo{pages}{1--7}.
\newblock
\showeprint[arxiv]{1803.11175}


\bibitem[\protect\citeauthoryear{Cheng, Wang, Zhang, O’Connell, Gray, Harper,
  and Zhu}{Cheng et~al\mbox{.}}{2019}]%
        {cheng2019explaining}
\bibfield{author}{\bibinfo{person}{Hao-Fei Cheng}, \bibinfo{person}{Ruotong
  Wang}, \bibinfo{person}{Zheng Zhang}, \bibinfo{person}{Fiona O’Connell},
  \bibinfo{person}{Terrance Gray}, \bibinfo{person}{F.~Maxwell Harper}, {and}
  \bibinfo{person}{Haiyi Zhu}.} \bibinfo{year}{2019}\natexlab{}.
\newblock \showarticletitle{Explaining Decision-Making Algorithms through UI:
  Strategies to Help Non-Expert Stakeholders}. In
  \bibinfo{booktitle}{\emph{Proc. of the 2019 CHI Conference on Human Factors
  in Computing Systems}} \emph{(\bibinfo{series}{CHI ’19})}.
  \bibinfo{publisher}{ACM}, \bibinfo{address}{New York, NY, USA},
  \bibinfo{pages}{1–12}.
\newblock
\showISBNx{9781450359702}


\bibitem[\protect\citeauthoryear{Ciampaglia}{Ciampaglia}{2018}]%
        {ciampaglia2018fighting}
\bibfield{author}{\bibinfo{person}{Giovanni~Luca Ciampaglia}.}
  \bibinfo{year}{2018}\natexlab{}.
\newblock \showarticletitle{Fighting fake news: a role for computational social
  science in the fight against digital misinformation}.
\newblock \bibinfo{journal}{\emph{Journal of Computational Social Science}}
  \bibinfo{volume}{1}, \bibinfo{number}{1} (\bibinfo{year}{2018}),
  \bibinfo{pages}{147--153}.
\newblock


\bibitem[\protect\citeauthoryear{Cohen, Hamilton, and Turner}{Cohen
  et~al\mbox{.}}{2011}]%
        {cohen2011computational}
\bibfield{author}{\bibinfo{person}{Sarah Cohen}, \bibinfo{person}{James~T
  Hamilton}, {and} \bibinfo{person}{Fred Turner}.}
  \bibinfo{year}{2011}\natexlab{}.
\newblock \showarticletitle{Computational journalism}.
\newblock \bibinfo{journal}{\emph{Commun. ACM}} \bibinfo{volume}{54},
  \bibinfo{number}{10} (\bibinfo{year}{2011}), \bibinfo{pages}{66--71}.
\newblock


\bibitem[\protect\citeauthoryear{Dilworth}{Dilworth}{1987}]%
        {dilworth2009decomposition}
\bibfield{author}{\bibinfo{person}{Robert~P Dilworth}.}
  \bibinfo{year}{1987}\natexlab{}.
\newblock \bibinfo{booktitle}{\emph{A Decomposition Theorem for Partially
  Ordered Sets}}.
\newblock \bibinfo{publisher}{Birkh{\"a}user Boston}, \bibinfo{address}{Boston,
  MA}, \bibinfo{pages}{139--144}.
\newblock
\showISBNx{978-0-8176-4842-8}


\bibitem[\protect\citeauthoryear{Eades and Huang}{Eades and Huang}{2004}]%
        {eades2004navigating}
\bibfield{author}{\bibinfo{person}{Peter Eades} {and} \bibinfo{person}{Mao~Lin
  Huang}.} \bibinfo{year}{2004}\natexlab{}.
\newblock \showarticletitle{Navigating Clustered Graphs Using Force-Directed
  Methods}.
\newblock In \bibinfo{booktitle}{\emph{Graph Algorithms and Applications 2}}.
  \bibinfo{publisher}{World Scientific}, \bibinfo{address}{5 Toh Tuck Link,
  Singapore, 596224, Singapore}, \bibinfo{pages}{191--215}.
\newblock


\bibitem[\protect\citeauthoryear{Ellson, Gansner, Koutsofios, North, and
  Woodhull}{Ellson et~al\mbox{.}}{2004}]%
        {ellson2004graphviz}
\bibfield{author}{\bibinfo{person}{John Ellson}, \bibinfo{person}{Emden~R.
  Gansner}, \bibinfo{person}{Eleftherios Koutsofios},
  \bibinfo{person}{Stephen~C. North}, {and} \bibinfo{person}{Gordon Woodhull}.}
  \bibinfo{year}{2004}\natexlab{}.
\newblock \showarticletitle{Graphviz and Dynagraph --- Static and Dynamic Graph
  Drawing Tools}.
\newblock In \bibinfo{booktitle}{\emph{Graph Drawing Software}},
  \bibfield{editor}{\bibinfo{person}{Michael J{\"u}nger} {and}
  \bibinfo{person}{Petra Mutzel}} (Eds.). \bibinfo{publisher}{Springer Berlin
  Heidelberg}, \bibinfo{address}{Berlin, Heidelberg},
  \bibinfo{pages}{127--148}.
\newblock
\showISBNx{978-3-642-18638-7}


\bibitem[\protect\citeauthoryear{Faloutsos, McCurley, and Tomkins}{Faloutsos
  et~al\mbox{.}}{2004}]%
        {faloutsos2004fast}
\bibfield{author}{\bibinfo{person}{Christos Faloutsos},
  \bibinfo{person}{Kevin~S. McCurley}, {and} \bibinfo{person}{Andrew Tomkins}.}
  \bibinfo{year}{2004}\natexlab{}.
\newblock \showarticletitle{Fast Discovery of Connection Subgraphs}. In
  \bibinfo{booktitle}{\emph{Proc. of the Tenth ACM SIGKDD Int. Conf. on
  Knowledge Discovery and Data Mining}} \emph{(\bibinfo{series}{KDD ’04})}.
  \bibinfo{publisher}{ACM}, \bibinfo{address}{New York, NY, USA},
  \bibinfo{pages}{118–127}.
\newblock
\showISBNx{1581138881}


\bibitem[\protect\citeauthoryear{Forsyth, Bittner, Musick, and Bello}{Forsyth
  et~al\mbox{.}}{2015}]%
        {forsyth2015improving}
\bibfield{author}{\bibinfo{person}{{Jonathan P.} Forsyth},
  \bibinfo{person}{{Rachel M.} Bittner}, \bibinfo{person}{Michael Musick},
  {and} \bibinfo{person}{{Juan P.} Bello}.} \bibinfo{year}{2015}\natexlab{}.
\newblock \showarticletitle{Improving and adapting finite state transducer
  methods for musical accompaniment}. In \bibinfo{booktitle}{\emph{41st Int.
  Computer Music Conf., ICMC 2015}}, \bibfield{editor}{\bibinfo{person}{Richard
  Dudas}} (Ed.). \bibinfo{publisher}{Int. Computer Music Association},
  \bibinfo{address}{Denton, TX}, \bibinfo{pages}{290--297}.
\newblock


\bibitem[\protect\citeauthoryear{Gansner, Koren, and North}{Gansner
  et~al\mbox{.}}{2005}]%
        {gansner2005topological}
\bibfield{author}{\bibinfo{person}{Emden~R Gansner}, \bibinfo{person}{Yehuda
  Koren}, {and} \bibinfo{person}{Stephen~C North}.}
  \bibinfo{year}{2005}\natexlab{}.
\newblock \showarticletitle{Topological fisheye views for visualizing large
  graphs}.
\newblock \bibinfo{journal}{\emph{IEEE Transactions on Visualization and
  Computer Graphics}} \bibinfo{volume}{11}, \bibinfo{number}{4}
  (\bibinfo{year}{2005}), \bibinfo{pages}{457--468}.
\newblock


\bibitem[\protect\citeauthoryear{Gansner, Koutsofios, North, and Vo}{Gansner
  et~al\mbox{.}}{1993}]%
        {gansner1993technique}
\bibfield{author}{\bibinfo{person}{Emden~R Gansner},
  \bibinfo{person}{Eleftherios Koutsofios}, \bibinfo{person}{Stephen~C North},
  {and} \bibinfo{person}{K-P Vo}.} \bibinfo{year}{1993}\natexlab{}.
\newblock \showarticletitle{A technique for drawing directed graphs}.
\newblock \bibinfo{journal}{\emph{IEEE Transactions on Software Engineering}}
  \bibinfo{volume}{19}, \bibinfo{number}{3} (\bibinfo{year}{1993}),
  \bibinfo{pages}{214--230}.
\newblock


\bibitem[\protect\citeauthoryear{Garc{\'\i}a, Moraga, Serrano, and
  Piattini}{Garc{\'\i}a et~al\mbox{.}}{2015}]%
        {garcia2015visualisation}
\bibfield{author}{\bibinfo{person}{F{\'e}lix Garc{\'\i}a},
  \bibinfo{person}{M{\textordfeminine}~{\'A}ngeles Moraga},
  \bibinfo{person}{Manuel Serrano}, {and} \bibinfo{person}{Mario Piattini}.}
  \bibinfo{year}{2015}\natexlab{}.
\newblock \showarticletitle{Visualisation environment for global software
  development management}.
\newblock \bibinfo{journal}{\emph{IET Software}} \bibinfo{volume}{9},
  \bibinfo{number}{2} (\bibinfo{year}{2015}), \bibinfo{pages}{51--64}.
\newblock


\bibitem[\protect\citeauthoryear{Genette}{Genette}{1983}]%
        {genette1983narrative}
\bibfield{author}{\bibinfo{person}{G{\'e}rard Genette}.}
  \bibinfo{year}{1983}\natexlab{}.
\newblock \bibinfo{booktitle}{\emph{Narrative discourse: An essay in method}}.
  Vol.~\bibinfo{volume}{3}.
\newblock \bibinfo{publisher}{Cornell University Press}, \bibinfo{address}{512
  East State St., Ithaca, NY}.
\newblock


\bibitem[\protect\citeauthoryear{Gleibs}{Gleibs}{2017}]%
        {gleibs2017all}
\bibfield{author}{\bibinfo{person}{Ilka~H Gleibs}.}
  \bibinfo{year}{2017}\natexlab{}.
\newblock \showarticletitle{Are all “research fields” equal? Rethinking
  practice for the use of data from crowdsourcing market places}.
\newblock \bibinfo{journal}{\emph{Behavior Research Methods}}
  \bibinfo{volume}{49}, \bibinfo{number}{4} (\bibinfo{year}{2017}),
  \bibinfo{pages}{1333--1342}.
\newblock


\bibitem[\protect\citeauthoryear{Hadjila, Belabed, and Merzoug}{Hadjila
  et~al\mbox{.}}{2019}]%
        {hadjila2019flexible}
\bibfield{author}{\bibinfo{person}{Fethallah Hadjila}, \bibinfo{person}{Amine
  Belabed}, {and} \bibinfo{person}{Mohammed Merzoug}.}
  \bibinfo{year}{2019}\natexlab{}.
\newblock \showarticletitle{Flexible service discovery based on multiple
  matching algorithms}.
\newblock \bibinfo{journal}{\emph{Int. Journal of Web Engineering and
  Technology}} \bibinfo{volume}{14}, \bibinfo{number}{4}
  (\bibinfo{year}{2019}), \bibinfo{pages}{315--340}.
\newblock


\bibitem[\protect\citeauthoryear{Hagberg, Swart, and S~Chult}{Hagberg
  et~al\mbox{.}}{2008}]%
        {hagberg2008exploring}
\bibfield{author}{\bibinfo{person}{Aric Hagberg}, \bibinfo{person}{Pieter
  Swart}, {and} \bibinfo{person}{Daniel S~Chult}.}
  \bibinfo{year}{2008}\natexlab{}.
\newblock \bibinfo{booktitle}{\emph{Exploring network structure, dynamics, and
  function using NetworkX}}.
\newblock \bibinfo{type}{{T}echnical {R}eport}. \bibinfo{institution}{Los
  Alamos National Lab. (LANL)}.
\newblock


\bibitem[\protect\citeauthoryear{Halverson, Corman, and Goodall}{Halverson
  et~al\mbox{.}}{2011}]%
        {halverson2011master}
\bibfield{author}{\bibinfo{person}{Jeffry Halverson}, \bibinfo{person}{Steven
  Corman}, {and} \bibinfo{person}{H~Lloyd Goodall}.}
  \bibinfo{year}{2011}\natexlab{}.
\newblock \bibinfo{booktitle}{\emph{Master narratives of Islamist extremism}}.
\newblock \bibinfo{publisher}{Springer}, \bibinfo{address}{175 5th Ave., New
  York, NY, USA}.
\newblock


\bibitem[\protect\citeauthoryear{Hauser, Paolacci, and Chandler}{Hauser
  et~al\mbox{.}}{2018}]%
        {hauser2018common}
\bibfield{author}{\bibinfo{person}{David Hauser}, \bibinfo{person}{Gabriele
  Paolacci}, {and} \bibinfo{person}{Jesse~J Chandler}.}
  \bibinfo{year}{2018}\natexlab{}.
\newblock \bibinfo{title}{Common Concerns with MTurk as a Participant Pool:
  Evidence and Solutions}.
\newblock
\newblock


\bibitem[\protect\citeauthoryear{Ho and Tang}{Ho and Tang}{2001}]%
        {ho2001towards}
\bibfield{author}{\bibinfo{person}{Jinwon Ho} {and} \bibinfo{person}{Rong
  Tang}.} \bibinfo{year}{2001}\natexlab{}.
\newblock \showarticletitle{Towards an Optimal Resolution to Information
  Overload: An Infomediary Approach}. In \bibinfo{booktitle}{\emph{Proc. of the
  2001 Int. ACM SIGGROUP Conference on Supporting Group Work}}
  \emph{(\bibinfo{series}{GROUP ’01})}. \bibinfo{publisher}{ACM},
  \bibinfo{address}{New York, NY, USA}, \bibinfo{pages}{91–96}.
\newblock
\showISBNx{1581132948}


\bibitem[\protect\citeauthoryear{Husfeldt}{Husfeldt}{1995}]%
        {husfeldt1995fully}
\bibfield{author}{\bibinfo{person}{Thore Husfeldt}.}
  \bibinfo{year}{1995}\natexlab{}.
\newblock \showarticletitle{Fully Dynamic Transitive Closure in plane dags with
  one source and one sink}. In \bibinfo{booktitle}{\emph{Algorithms --- ESA
  '95}}, \bibfield{editor}{\bibinfo{person}{Paul Spirakis}} (Ed.).
  \bibinfo{publisher}{Springer Berlin Heidelberg}, \bibinfo{address}{Berlin,
  Heidelberg}, \bibinfo{pages}{199--212}.
\newblock
\showISBNx{978-3-540-44913-3}


\bibitem[\protect\citeauthoryear{Kim and Oh}{Kim and Oh}{2011}]%
        {kim2011topic}
\bibfield{author}{\bibinfo{person}{Dongwoo Kim} {and} \bibinfo{person}{Alice
  Oh}.} \bibinfo{year}{2011}\natexlab{}.
\newblock \showarticletitle{Topic Chains for Understanding a News Corpus}. In
  \bibinfo{booktitle}{\emph{Computational Linguistics and Intelligent Text
  Processing}}, \bibfield{editor}{\bibinfo{person}{Alexander Gelbukh}} (Ed.).
  \bibinfo{publisher}{Springer Berlin Heidelberg}, \bibinfo{address}{Berlin,
  Heidelberg}, \bibinfo{pages}{163--176}.
\newblock
\showISBNx{978-3-642-19437-5}


\bibitem[\protect\citeauthoryear{Kohler~Riessman}{Kohler~Riessman}{2000}]%
        {kohler2000analysis}
\bibfield{author}{\bibinfo{person}{Catherine Kohler~Riessman}.}
  \bibinfo{year}{2000}\natexlab{}.
\newblock \showarticletitle{Analysis of personal narratives}.
\newblock In \bibinfo{booktitle}{\emph{Qualitative research in social work}}.
  \bibinfo{publisher}{Columbia University Press}, \bibinfo{address}{61 West 62
  St., New York, NY, USA}, Chapter~7, \bibinfo{pages}{168--191}.
\newblock


\bibitem[\protect\citeauthoryear{Lasecki, Gordon, Koutra, Jung, Dow, and
  Bigham}{Lasecki et~al\mbox{.}}{2014}]%
        {lasecki2014glance}
\bibfield{author}{\bibinfo{person}{Walter~S. Lasecki},
  \bibinfo{person}{Mitchell Gordon}, \bibinfo{person}{Danai Koutra},
  \bibinfo{person}{Malte~F. Jung}, \bibinfo{person}{Steven~P. Dow}, {and}
  \bibinfo{person}{Jeffrey~P. Bigham}.} \bibinfo{year}{2014}\natexlab{}.
\newblock \showarticletitle{Glance: Rapidly Coding Behavioral Video with the
  Crowd}. In \bibinfo{booktitle}{\emph{Proc. of the 27th Annual ACM Symposium
  on User Interface Software and Technology}} \emph{(\bibinfo{series}{UIST
  ’14})}. \bibinfo{publisher}{ACM}, \bibinfo{address}{New York, NY, USA},
  \bibinfo{pages}{551–562}.
\newblock
\showISBNx{9781450330695}


\bibitem[\protect\citeauthoryear{Li, Manns, North, and Luther}{Li
  et~al\mbox{.}}{2019}]%
        {li2019dropping}
\bibfield{author}{\bibinfo{person}{Tianyi Li}, \bibinfo{person}{Chandler~J
  Manns}, \bibinfo{person}{Chris North}, {and} \bibinfo{person}{Kurt Luther}.}
  \bibinfo{year}{2019}\natexlab{}.
\newblock \showarticletitle{Dropping the baton? Understanding errors and
  bottlenecks in a crowdsourced sensemaking pipeline}.
\newblock \bibinfo{journal}{\emph{Proceedings of the ACM on Human-Computer
  Interaction}} \bibinfo{volume}{3}, \bibinfo{number}{CSCW}
  (\bibinfo{year}{2019}), \bibinfo{pages}{1--26}.
\newblock


\bibitem[\protect\citeauthoryear{Li, Shah, Luther, and North}{Li
  et~al\mbox{.}}{2018}]%
        {li2018crowdsourcing}
\bibfield{author}{\bibinfo{person}{Tianyi Li}, \bibinfo{person}{Asmita Shah},
  \bibinfo{person}{Kurt Luther}, {and} \bibinfo{person}{Chris North}.}
  \bibinfo{year}{2018}\natexlab{}.
\newblock \showarticletitle{Crowdsourcing Intelligence Analysis with Context
  Slices}. In \bibinfo{booktitle}{\emph{CHI 2018 Sensemaking Workshop}}.
\newblock


\bibitem[\protect\citeauthoryear{Liu, Niu, Lai, Kong, and Xu}{Liu
  et~al\mbox{.}}{2017}]%
        {liu2017growing}
\bibfield{author}{\bibinfo{person}{Bang Liu}, \bibinfo{person}{Di Niu},
  \bibinfo{person}{Kunfeng Lai}, \bibinfo{person}{Linglong Kong}, {and}
  \bibinfo{person}{Yu Xu}.} \bibinfo{year}{2017}\natexlab{}.
\newblock \showarticletitle{Growing Story Forest Online from Massive Breaking
  News}. In \bibinfo{booktitle}{\emph{Proc. of the 2017 ACM on Conference on
  Information and Knowledge Management}} \emph{(\bibinfo{series}{CIKM ’17})}.
  \bibinfo{publisher}{ACM}, \bibinfo{address}{New York, NY, USA},
  \bibinfo{pages}{777–785}.
\newblock
\showISBNx{9781450349185}


\bibitem[\protect\citeauthoryear{McInnes, Healy, and Astels}{McInnes
  et~al\mbox{.}}{2017}]%
        {mcinnes2017hdbscan}
\bibfield{author}{\bibinfo{person}{Leland McInnes}, \bibinfo{person}{John
  Healy}, {and} \bibinfo{person}{Steve Astels}.}
  \bibinfo{year}{2017}\natexlab{}.
\newblock \showarticletitle{hdbscan: Hierarchical density based clustering}.
\newblock \bibinfo{journal}{\emph{Journal of Open Source Software}}
  \bibinfo{volume}{2}, \bibinfo{number}{11} (\bibinfo{year}{2017}),
  \bibinfo{pages}{205}.
\newblock


\bibitem[\protect\citeauthoryear{McInnes, Healy, and Melville}{McInnes
  et~al\mbox{.}}{2018}]%
        {mcinnes2018umap}
\bibfield{author}{\bibinfo{person}{Leland McInnes}, \bibinfo{person}{John
  Healy}, {and} \bibinfo{person}{James Melville}.}
  \bibinfo{year}{2018}\natexlab{}.
\newblock \bibinfo{title}{Umap: Uniform manifold approximation and projection
  for dimension reduction}.
\newblock \bibinfo{howpublished}{arXiv preprint arXiv:1802.03426}.
\newblock


\bibitem[\protect\citeauthoryear{Milo, Shen-Orr, Itzkovitz, Kashtan,
  Chklovskii, and Alon}{Milo et~al\mbox{.}}{2002}]%
        {milo2002network}
\bibfield{author}{\bibinfo{person}{Ron Milo}, \bibinfo{person}{Shai Shen-Orr},
  \bibinfo{person}{Shalev Itzkovitz}, \bibinfo{person}{Nadav Kashtan},
  \bibinfo{person}{Dmitri Chklovskii}, {and} \bibinfo{person}{Uri Alon}.}
  \bibinfo{year}{2002}\natexlab{}.
\newblock \showarticletitle{Network motifs: simple building blocks of complex
  networks}.
\newblock \bibinfo{journal}{\emph{Science}} \bibinfo{volume}{298},
  \bibinfo{number}{5594} (\bibinfo{year}{2002}), \bibinfo{pages}{824--827}.
\newblock


\bibitem[\protect\citeauthoryear{Miskimmon, O'loughlin, and Roselle}{Miskimmon
  et~al\mbox{.}}{2014}]%
        {miskimmon2014strategic}
\bibfield{author}{\bibinfo{person}{Alister Miskimmon}, \bibinfo{person}{Ben
  O'loughlin}, {and} \bibinfo{person}{Laura Roselle}.}
  \bibinfo{year}{2014}\natexlab{}.
\newblock \bibinfo{booktitle}{\emph{Strategic narratives: Communication power
  and the new world order}}.
\newblock \bibinfo{publisher}{Routledge}, \bibinfo{address}{711 3rd Ave. \#8,
  New York, NY, USA}.
\newblock


\bibitem[\protect\citeauthoryear{Miskimmon, O'Loughlin, and Roselle}{Miskimmon
  et~al\mbox{.}}{2017}]%
        {miskimmon2017forging}
\bibfield{author}{\bibinfo{person}{Alister Miskimmon}, \bibinfo{person}{Ben
  O'Loughlin}, {and} \bibinfo{person}{Laura Roselle}.}
  \bibinfo{year}{2017}\natexlab{}.
\newblock \bibinfo{booktitle}{\emph{Forging the world: Strategic narratives and
  international relations}}.
\newblock \bibinfo{publisher}{University of Michigan Press},
  \bibinfo{address}{839 Greene St., Ann Arbor, MI}.
\newblock


\bibitem[\protect\citeauthoryear{Mitra and Gilbert}{Mitra and Gilbert}{2015}]%
        {mitra2015credbank}
\bibfield{author}{\bibinfo{person}{Tanushree Mitra} {and} \bibinfo{person}{Eric
  Gilbert}.} \bibinfo{year}{2015}\natexlab{}.
\newblock \showarticletitle{CREDBANK: A Large-Scale Social Media Corpus With
  Associated Credibility Annotations}.
\newblock \bibinfo{journal}{\emph{Int. AAAI Conf. on Web and Social Media}}
  \bibinfo{volume}{2015} (\bibinfo{year}{2015}), \bibinfo{pages}{258--267}.
\newblock


\bibitem[\protect\citeauthoryear{Mitra, Hutto, and Gilbert}{Mitra
  et~al\mbox{.}}{2015}]%
        {mitra2015comparing}
\bibfield{author}{\bibinfo{person}{Tanushree Mitra}, \bibinfo{person}{C.J.
  Hutto}, {and} \bibinfo{person}{Eric Gilbert}.}
  \bibinfo{year}{2015}\natexlab{}.
\newblock \showarticletitle{Comparing Person- and Process-Centric Strategies
  for Obtaining Quality Data on Amazon Mechanical Turk}. In
  \bibinfo{booktitle}{\emph{Proc. of the 33rd Annual ACM Conference on Human
  Factors in Computing Systems}} \emph{(\bibinfo{series}{CHI ’15})}.
  \bibinfo{publisher}{ACM}, \bibinfo{address}{New York, NY, USA},
  \bibinfo{pages}{1345–1354}.
\newblock
\showISBNx{9781450331456}


\bibitem[\protect\citeauthoryear{Nallapati, Feng, Peng, and Allan}{Nallapati
  et~al\mbox{.}}{2004}]%
        {nallapati2004event}
\bibfield{author}{\bibinfo{person}{Ramesh Nallapati}, \bibinfo{person}{Ao
  Feng}, \bibinfo{person}{Fuchun Peng}, {and} \bibinfo{person}{James Allan}.}
  \bibinfo{year}{2004}\natexlab{}.
\newblock \showarticletitle{Event Threading within News Topics}. In
  \bibinfo{booktitle}{\emph{Proc. of the Thirteenth ACM Int. Conf. on
  Information and Knowledge Management}} \emph{(\bibinfo{series}{CIKM ’04})}.
  \bibinfo{publisher}{ACM}, \bibinfo{address}{New York, NY, USA},
  \bibinfo{pages}{446–453}.
\newblock
\showISBNx{1581138741}


\bibitem[\protect\citeauthoryear{Neigel, Caylor, Kase, Vanni, and Hoye}{Neigel
  et~al\mbox{.}}{2018}]%
        {neigel2018role}
\bibfield{author}{\bibinfo{person}{Alexis~R Neigel}, \bibinfo{person}{Justine~P
  Caylor}, \bibinfo{person}{Sue~E Kase}, \bibinfo{person}{Michelle~T Vanni},
  {and} \bibinfo{person}{Jefferson Hoye}.} \bibinfo{year}{2018}\natexlab{}.
\newblock \showarticletitle{The role of trust and automation in an intelligence
  analyst decisional guidance paradigm}.
\newblock \bibinfo{journal}{\emph{Journal of Cognitive Engineering and Decision
  Making}} \bibinfo{volume}{12}, \bibinfo{number}{4} (\bibinfo{year}{2018}),
  \bibinfo{pages}{239--247}.
\newblock


\bibitem[\protect\citeauthoryear{Nied, Stewart, Spiro, and Starbird}{Nied
  et~al\mbox{.}}{2017}]%
        {nied2017alternative}
\bibfield{author}{\bibinfo{person}{A.~Conrad Nied}, \bibinfo{person}{Leo
  Stewart}, \bibinfo{person}{Emma Spiro}, {and} \bibinfo{person}{Kate
  Starbird}.} \bibinfo{year}{2017}\natexlab{}.
\newblock \showarticletitle{Alternative Narratives of Crisis Events:
  Communities and Social Botnets Engaged on Social Media}. In
  \bibinfo{booktitle}{\emph{Companion of the 2017 ACM Conference on Computer
  Supported Cooperative Work and Social Computing}}
  \emph{(\bibinfo{series}{CSCW ’17 Companion})}. \bibinfo{publisher}{ACM},
  \bibinfo{address}{New York, NY, USA}, \bibinfo{pages}{263–266}.
\newblock
\showISBNx{9781450346887}


\bibitem[\protect\citeauthoryear{Norambuena, Horning, and Mitra}{Norambuena
  et~al\mbox{.}}{2020}]%
        {norambuenaevaluating}
\bibfield{author}{\bibinfo{person}{Brian~Keith Norambuena},
  \bibinfo{person}{Michael Horning}, {and} \bibinfo{person}{Tanushree Mitra}.}
  \bibinfo{year}{2020}\natexlab{}.
\newblock \showarticletitle{Evaluating the Inverted Pyramid Structure through
  Automatic 5W1H Extraction and Summarization}. In
  \bibinfo{booktitle}{\emph{Proc. of the 2020 Computation + Journalism
  Symposium}}. \bibinfo{publisher}{Computation + Journalism 2020},
  \bibinfo{address}{Boston, MA, USA}, \bibinfo{pages}{1--7}.
\newblock


\bibitem[\protect\citeauthoryear{Paolacci, Chandler, and Ipeirotis}{Paolacci
  et~al\mbox{.}}{2010}]%
        {paolacci2010running}
\bibfield{author}{\bibinfo{person}{Gabriele Paolacci}, \bibinfo{person}{Jesse
  Chandler}, {and} \bibinfo{person}{Panagiotis~G Ipeirotis}.}
  \bibinfo{year}{2010}\natexlab{}.
\newblock \showarticletitle{Running experiments on amazon mechanical turk}.
\newblock \bibinfo{journal}{\emph{Judgment and Decision making}}
  \bibinfo{volume}{5}, \bibinfo{number}{5} (\bibinfo{year}{2010}),
  \bibinfo{pages}{411--419}.
\newblock


\bibitem[\protect\citeauthoryear{Pijls and Potharst}{Pijls and
  Potharst}{2013}]%
        {pijls2013another}
\bibfield{author}{\bibinfo{person}{Wim Pijls} {and} \bibinfo{person}{Rob
  Potharst}.} \bibinfo{year}{2013}\natexlab{}.
\newblock \showarticletitle{Another Note on Dilworth's Decomposition Theorem}.
\newblock \bibinfo{journal}{\emph{Journal of Discrete Mathematics}}
  \bibinfo{volume}{2013} (\bibinfo{year}{2013}), \bibinfo{pages}{692--645}.
\newblock
\showISSN{2090-9837}


\bibitem[\protect\citeauthoryear{Puckett}{Puckett}{2016}]%
        {puckett2016narrative}
\bibfield{author}{\bibinfo{person}{Kent Puckett}.}
  \bibinfo{year}{2016}\natexlab{}.
\newblock \bibinfo{booktitle}{\emph{Narrative theory}}.
\newblock \bibinfo{publisher}{Cambridge University Press},
  \bibinfo{address}{One Liberty Plaza, New York, NY, USA}.
\newblock


\bibitem[\protect\citeauthoryear{Rimmon-Kenan}{Rimmon-Kenan}{2003}]%
        {rimmon2003narrative}
\bibfield{author}{\bibinfo{person}{Shlomith Rimmon-Kenan}.}
  \bibinfo{year}{2003}\natexlab{}.
\newblock \bibinfo{booktitle}{\emph{Narrative fiction: Contemporary poetics}}.
\newblock \bibinfo{publisher}{Routledge}, \bibinfo{address}{711 3rd Ave. \#8,
  New York, NY, USA}.
\newblock


\bibitem[\protect\citeauthoryear{Robbins}{Robbins}{2019}]%
        {robbins2019weaponization}
\bibfield{author}{\bibinfo{person}{Corina Robbins}.}
  \bibinfo{year}{2019}\natexlab{}.
\newblock \bibinfo{booktitle}{\emph{The Weaponization of Social Media}}.
\newblock \bibinfo{type}{{T}echnical {R}eport}. \bibinfo{institution}{Mercy
  Corps}.
\newblock


\bibitem[\protect\citeauthoryear{Roselle, Miskimmon, and O’Loughlin}{Roselle
  et~al\mbox{.}}{2014}]%
        {roselle2014strategic}
\bibfield{author}{\bibinfo{person}{Laura Roselle}, \bibinfo{person}{Alister
  Miskimmon}, {and} \bibinfo{person}{Ben O’Loughlin}.}
  \bibinfo{year}{2014}\natexlab{}.
\newblock \showarticletitle{Strategic narrative: A new means to understand soft
  power}.
\newblock \bibinfo{journal}{\emph{Media, War \& Conflict}} \bibinfo{volume}{7},
  \bibinfo{number}{1} (\bibinfo{year}{2014}), \bibinfo{pages}{70--84}.
\newblock


\bibitem[\protect\citeauthoryear{Salehi, Irani, Bernstein, Alkhatib, Ogbe,
  Milland, and Clickhappier}{Salehi et~al\mbox{.}}{2015}]%
        {salehi2015we}
\bibfield{author}{\bibinfo{person}{Niloufar Salehi}, \bibinfo{person}{Lilly~C.
  Irani}, \bibinfo{person}{Michael~S. Bernstein}, \bibinfo{person}{Ali
  Alkhatib}, \bibinfo{person}{Eva Ogbe}, \bibinfo{person}{Kristy Milland},
  {and} \bibinfo{person}{Clickhappier}.} \bibinfo{year}{2015}\natexlab{}.
\newblock \showarticletitle{We Are Dynamo: Overcoming Stalling and Friction in
  Collective Action for Crowd Workers}. In \bibinfo{booktitle}{\emph{Proc. of
  the 33rd Annual ACM Conference on Human Factors in Computing Systems}}
  \emph{(\bibinfo{series}{CHI '15})}. \bibinfo{publisher}{ACM},
  \bibinfo{address}{New York, NY, USA}, \bibinfo{pages}{1621–1630}.
\newblock
\showISBNx{9781450331456}


\bibitem[\protect\citeauthoryear{Shahaf and Guestrin}{Shahaf and
  Guestrin}{2010}]%
        {shahaf2010connecting}
\bibfield{author}{\bibinfo{person}{Dafna Shahaf} {and} \bibinfo{person}{Carlos
  Guestrin}.} \bibinfo{year}{2010}\natexlab{}.
\newblock \showarticletitle{Connecting the Dots between News Articles}. In
  \bibinfo{booktitle}{\emph{Proc. of the 16th ACM SIGKDD Int. Conf. on
  Knowledge Discovery and Data Mining}} \emph{(\bibinfo{series}{KDD ’10})}.
  \bibinfo{publisher}{ACM}, \bibinfo{address}{New York, NY, USA},
  \bibinfo{pages}{623–632}.
\newblock
\showISBNx{9781450300551}


\bibitem[\protect\citeauthoryear{Shahaf and Guestrin}{Shahaf and
  Guestrin}{2012}]%
        {shahaf2012connecting}
\bibfield{author}{\bibinfo{person}{Dafna Shahaf} {and} \bibinfo{person}{Carlos
  Guestrin}.} \bibinfo{year}{2012}\natexlab{}.
\newblock \showarticletitle{Connecting two (or less) dots: Discovering
  structure in news articles}.
\newblock \bibinfo{journal}{\emph{ACM Transactions on Knowledge Discovery from
  Data (TKDD)}} \bibinfo{volume}{5}, \bibinfo{number}{4}
  (\bibinfo{year}{2012}), \bibinfo{pages}{1--31}.
\newblock


\bibitem[\protect\citeauthoryear{Shahaf, Guestrin, and Horvitz}{Shahaf
  et~al\mbox{.}}{2012}]%
        {shahaf2012trains}
\bibfield{author}{\bibinfo{person}{Dafna Shahaf}, \bibinfo{person}{Carlos
  Guestrin}, {and} \bibinfo{person}{Eric Horvitz}.}
  \bibinfo{year}{2012}\natexlab{}.
\newblock \showarticletitle{Trains of Thought: Generating Information Maps}. In
  \bibinfo{booktitle}{\emph{Proc. of the 21st Int. Conf. on World Wide Web}}
  \emph{(\bibinfo{series}{WWW ’12})}. \bibinfo{publisher}{ACM},
  \bibinfo{address}{New York, NY, USA}, \bibinfo{pages}{899–908}.
\newblock
\showISBNx{9781450312295}


\bibitem[\protect\citeauthoryear{Shahaf, Guestrin, and Horvitz}{Shahaf
  et~al\mbox{.}}{2013a}]%
        {shahaf2013metro}
\bibfield{author}{\bibinfo{person}{Dafna Shahaf}, \bibinfo{person}{Carlos
  Guestrin}, {and} \bibinfo{person}{Eric Horvitz}.}
  \bibinfo{year}{2013}\natexlab{a}.
\newblock \showarticletitle{Metro Maps of Information}.
\newblock \bibinfo{journal}{\emph{SIGWEB Newsl.}}  \bibinfo{volume}{Spring},
  Article \bibinfo{articleno}{4} (\bibinfo{year}{2013}),
  \bibinfo{numpages}{9}~pages.
\newblock
\showISSN{1931-1745}


\bibitem[\protect\citeauthoryear{Shahaf, Guestrin, Horvitz, and
  Leskovec}{Shahaf et~al\mbox{.}}{2015}]%
        {shahaf2015information}
\bibfield{author}{\bibinfo{person}{Dafna Shahaf}, \bibinfo{person}{Carlos
  Guestrin}, \bibinfo{person}{Eric Horvitz}, {and} \bibinfo{person}{Jure
  Leskovec}.} \bibinfo{year}{2015}\natexlab{}.
\newblock \showarticletitle{Information cartography}.
\newblock \bibinfo{journal}{\emph{Commun. ACM}} \bibinfo{volume}{58},
  \bibinfo{number}{11} (\bibinfo{year}{2015}), \bibinfo{pages}{62--73}.
\newblock


\bibitem[\protect\citeauthoryear{Shahaf, Yang, Suen, Jacobs, Wang, and
  Leskovec}{Shahaf et~al\mbox{.}}{2013b}]%
        {shahaf2013information}
\bibfield{author}{\bibinfo{person}{Dafna Shahaf}, \bibinfo{person}{Jaewon
  Yang}, \bibinfo{person}{Caroline Suen}, \bibinfo{person}{Jeff Jacobs},
  \bibinfo{person}{Heidi Wang}, {and} \bibinfo{person}{Jure Leskovec}.}
  \bibinfo{year}{2013}\natexlab{b}.
\newblock \showarticletitle{Information Cartography: Creating Zoomable,
  Large-Scale Maps of Information}. In \bibinfo{booktitle}{\emph{Proc. of the
  19th ACM SIGKDD Int. Conf. on Knowledge Discovery and Data Mining}}
  \emph{(\bibinfo{series}{KDD ’13})}. \bibinfo{publisher}{ACM},
  \bibinfo{address}{New York, NY, USA}, \bibinfo{pages}{1097–1105}.
\newblock
\showISBNx{9781450321747}


\bibitem[\protect\citeauthoryear{Shamim, Balakrishnan, and Tahir}{Shamim
  et~al\mbox{.}}{2015}]%
        {shamim2015evaluation}
\bibfield{author}{\bibinfo{person}{Azra Shamim}, \bibinfo{person}{Vimala
  Balakrishnan}, {and} \bibinfo{person}{Muhammad Tahir}.}
  \bibinfo{year}{2015}\natexlab{}.
\newblock \showarticletitle{Evaluation of opinion visualization techniques}.
\newblock \bibinfo{journal}{\emph{Information visualization}}
  \bibinfo{volume}{14}, \bibinfo{number}{4} (\bibinfo{year}{2015}),
  \bibinfo{pages}{339--358}.
\newblock


\bibitem[\protect\citeauthoryear{Smith}{Smith}{1980}]%
        {smith1980narrative}
\bibfield{author}{\bibinfo{person}{Barbara~Herrnstein Smith}.}
  \bibinfo{year}{1980}\natexlab{}.
\newblock \showarticletitle{Narrative versions, narrative theories}.
\newblock \bibinfo{journal}{\emph{Critical inquiry}} \bibinfo{volume}{7},
  \bibinfo{number}{1} (\bibinfo{year}{1980}), \bibinfo{pages}{213--236}.
\newblock


\bibitem[\protect\citeauthoryear{Soni, Mitra, Gilbert, and Eisenstein}{Soni
  et~al\mbox{.}}{2014}]%
        {soni2014modeling}
\bibfield{author}{\bibinfo{person}{Sandeep Soni}, \bibinfo{person}{Tanushree
  Mitra}, \bibinfo{person}{Eric Gilbert}, {and} \bibinfo{person}{Jacob
  Eisenstein}.} \bibinfo{year}{2014}\natexlab{}.
\newblock \showarticletitle{Modeling Factuality Judgments in Social Media
  Text}. In \bibinfo{booktitle}{\emph{Proc. of the 52nd Annual Meeting of the
  ACL (Volume 2: Short Papers)}}. \bibinfo{publisher}{ACL},
  \bibinfo{address}{Baltimore, Maryland}, \bibinfo{pages}{415--420}.
\newblock


\bibitem[\protect\citeauthoryear{Szostek}{Szostek}{2017}]%
        {szostek2017defence}
\bibfield{author}{\bibinfo{person}{Joanna Szostek}.}
  \bibinfo{year}{2017}\natexlab{}.
\newblock \showarticletitle{Defence and Promotion of Desired State Identity in
  Russia’s Strategic Narrative}.
\newblock \bibinfo{journal}{\emph{Geopolitics}} \bibinfo{volume}{22},
  \bibinfo{number}{3} (\bibinfo{year}{2017}), \bibinfo{pages}{571--593}.
\newblock


\bibitem[\protect\citeauthoryear{Tversky}{Tversky}{2004}]%
        {tversky2004narratives}
\bibfield{author}{\bibinfo{person}{Barbara Tversky}.}
  \bibinfo{year}{2004}\natexlab{}.
\newblock \showarticletitle{Narratives of Space, Time, and Life}.
\newblock \bibinfo{journal}{\emph{Mind \& Language}} \bibinfo{volume}{19},
  \bibinfo{number}{4} (\bibinfo{year}{2004}), \bibinfo{pages}{380--392}.
\newblock


\bibitem[\protect\citeauthoryear{Wang, Cardie, and Marchetti}{Wang
  et~al\mbox{.}}{2015}]%
        {wang2015socially}
\bibfield{author}{\bibinfo{person}{Lu Wang}, \bibinfo{person}{Claire Cardie},
  {and} \bibinfo{person}{Galen Marchetti}.} \bibinfo{year}{2015}\natexlab{}.
\newblock \showarticletitle{Socially-Informed Timeline Generation for Complex
  Events}. In \bibinfo{booktitle}{\emph{Proc. of the 2015 Conference of the
  North {A}merican Chapter of the ACL: Human Language Technologies}}.
  \bibinfo{publisher}{ACL}, \bibinfo{address}{Denver, Colorado},
  \bibinfo{pages}{1055--1065}.
\newblock


\bibitem[\protect\citeauthoryear{Weedon, Nuland, and Stamos}{Weedon
  et~al\mbox{.}}{2017}]%
        {weedon2017information}
\bibfield{author}{\bibinfo{person}{Jen Weedon}, \bibinfo{person}{William
  Nuland}, {and} \bibinfo{person}{Alex Stamos}.}
  \bibinfo{year}{2017}\natexlab{}.
\newblock \bibinfo{booktitle}{\emph{Information operations and Facebook}}.
\newblock \bibinfo{type}{{T}echnical {R}eport}.
  \bibinfo{institution}{Facebook}.
\newblock


\bibitem[\protect\citeauthoryear{Williamson}{Williamson}{2016}]%
        {williamson2016ethics}
\bibfield{author}{\bibinfo{person}{Vanessa Williamson}.}
  \bibinfo{year}{2016}\natexlab{}.
\newblock \showarticletitle{On the ethics of crowdsourced research}.
\newblock \bibinfo{journal}{\emph{PS: Political Science \& Politics}}
  \bibinfo{volume}{49}, \bibinfo{number}{1} (\bibinfo{year}{2016}),
  \bibinfo{pages}{77--81}.
\newblock


\bibitem[\protect\citeauthoryear{Wilson, Zhou, and Starbird}{Wilson
  et~al\mbox{.}}{2018}]%
        {wilson2018assembling}
\bibfield{author}{\bibinfo{person}{Tom Wilson}, \bibinfo{person}{Kaitlyn Zhou},
  {and} \bibinfo{person}{Kate Starbird}.} \bibinfo{year}{2018}\natexlab{}.
\newblock \showarticletitle{Assembling strategic narratives: Information
  operations as collaborative work within an online community}.
\newblock \bibinfo{journal}{\emph{Proc. of the ACM on HCI}}
  \bibinfo{volume}{2}, \bibinfo{number}{CSCW} (\bibinfo{year}{2018}),
  \bibinfo{pages}{1--26}.
\newblock


\bibitem[\protect\citeauthoryear{Worth}{Worth}{2008}]%
        {worth2008storytelling}
\bibfield{author}{\bibinfo{person}{Sarah~E. Worth}.}
  \bibinfo{year}{2008}\natexlab{}.
\newblock \showarticletitle{Storytelling and Narrative Knowing: An Examination
  of the Epistemic Benefits of Well-Told Stories}.
\newblock \bibinfo{journal}{\emph{Journal of Aesthetic Education}}
  \bibinfo{volume}{42}, \bibinfo{number}{3} (\bibinfo{year}{2008}),
  \bibinfo{pages}{42--56}.
\newblock
\showISSN{00218510, 15437809}


\bibitem[\protect\citeauthoryear{Xu, Wang, and Zhang}{Xu et~al\mbox{.}}{2013}]%
        {xu2013summarizing}
\bibfield{author}{\bibinfo{person}{Shize Xu}, \bibinfo{person}{Shanshan Wang},
  {and} \bibinfo{person}{Yan Zhang}.} \bibinfo{year}{2013}\natexlab{}.
\newblock \showarticletitle{Summarizing Complex Events: a Cross-Modal Solution
  of Storylines Extraction and Reconstruction}. In
  \bibinfo{booktitle}{\emph{Proc. of the 2013 Conference on Empirical Methods
  in Natural Language Processing}}. \bibinfo{publisher}{ACL},
  \bibinfo{address}{Seattle, Washington, USA}, \bibinfo{pages}{1281--1291}.
\newblock


\bibitem[\protect\citeauthoryear{Yan, Wan, Otterbacher, Kong, Li, and
  Zhang}{Yan et~al\mbox{.}}{2011}]%
        {yan2011evolutionary}
\bibfield{author}{\bibinfo{person}{Rui Yan}, \bibinfo{person}{Xiaojun Wan},
  \bibinfo{person}{Jahna Otterbacher}, \bibinfo{person}{Liang Kong},
  \bibinfo{person}{Xiaoming Li}, {and} \bibinfo{person}{Yan Zhang}.}
  \bibinfo{year}{2011}\natexlab{}.
\newblock \showarticletitle{Evolutionary Timeline Summarization: A Balanced
  Optimization Framework via Iterative Substitution}. In
  \bibinfo{booktitle}{\emph{Proc. of the 34th Int. ACM SIGIR Conference on
  Research and Development in Info. Retrieval}} \emph{(\bibinfo{series}{SIGIR
  ’11})}. \bibinfo{publisher}{ACM}, \bibinfo{address}{New York, NY, USA},
  \bibinfo{pages}{745–754}.
\newblock
\showISBNx{9781450307574}


\bibitem[\protect\citeauthoryear{Yang, Shi, and Wei}{Yang
  et~al\mbox{.}}{2009}]%
        {yang2009discovering}
\bibfield{author}{\bibinfo{person}{Christopher~C Yang},
  \bibinfo{person}{Xiaodong Shi}, {and} \bibinfo{person}{Chih-Ping Wei}.}
  \bibinfo{year}{2009}\natexlab{}.
\newblock \showarticletitle{Discovering event evolution graphs from news
  corpora}.
\newblock \bibinfo{journal}{\emph{IEEE Transactions on Systems, Man, and
  Cybernetics-Part A: Systems and Humans}} \bibinfo{volume}{39},
  \bibinfo{number}{4} (\bibinfo{year}{2009}), \bibinfo{pages}{850--863}.
\newblock


\bibitem[\protect\citeauthoryear{Zhang, Verou, and Karger}{Zhang
  et~al\mbox{.}}{2017}]%
        {zhang2017wikum}
\bibfield{author}{\bibinfo{person}{Amy~X. Zhang}, \bibinfo{person}{Lea Verou},
  {and} \bibinfo{person}{David Karger}.} \bibinfo{year}{2017}\natexlab{}.
\newblock \showarticletitle{Wikum: Bridging Discussion Forums and Wikis Using
  Recursive Summarization}. In \bibinfo{booktitle}{\emph{Proc. of the 2017 ACM
  Conference on Computer Supported Cooperative Work and Social Computing}}
  \emph{(\bibinfo{series}{CSCW ’17})}. \bibinfo{publisher}{ACM},
  \bibinfo{address}{New York, NY, USA}, \bibinfo{pages}{2082–2096}.
\newblock
\showISBNx{9781450343350}


\bibitem[\protect\citeauthoryear{Zhou, Yu, Hu, and Hu}{Zhou
  et~al\mbox{.}}{2017}]%
        {zhou2017survey}
\bibfield{author}{\bibinfo{person}{Houkui Zhou}, \bibinfo{person}{Huimin Yu},
  \bibinfo{person}{Roland Hu}, {and} \bibinfo{person}{Junguo Hu}.}
  \bibinfo{year}{2017}\natexlab{}.
\newblock \showarticletitle{A survey on trends of cross-media topic evolution
  map}.
\newblock \bibinfo{journal}{\emph{Knowledge-Based Systems}}
  \bibinfo{volume}{124} (\bibinfo{year}{2017}), \bibinfo{pages}{164--175}.
\newblock


\bibitem[\protect\citeauthoryear{{Zhou}, {Wu}, and {Cao}}{{Zhou}
  et~al\mbox{.}}{2017}]%
        {zhou2017emmbtt}
\bibfield{author}{\bibinfo{person}{P. {Zhou}}, \bibinfo{person}{B. {Wu}}, {and}
  \bibinfo{person}{Z. {Cao}}.} \bibinfo{year}{2017}\natexlab{}.
\newblock \showarticletitle{EMMBTT: A Novel Event Evolution Model Based on
  TFxIEF and TDC in Tracking News Streams}. In \bibinfo{booktitle}{\emph{2017
  IEEE Second Int. Conf. on Data Science in Cyberspace (DSC)}}.
  \bibinfo{publisher}{IEEE}, \bibinfo{address}{Shenzhen, China},
  \bibinfo{pages}{102--107}.
\newblock


\end{thebibliography}

\appendix
\newpage
\section{Mechanical Turk Job}
\begin{figure}[htbp]
    \centering
    \includegraphics[width=0.7\textwidth]{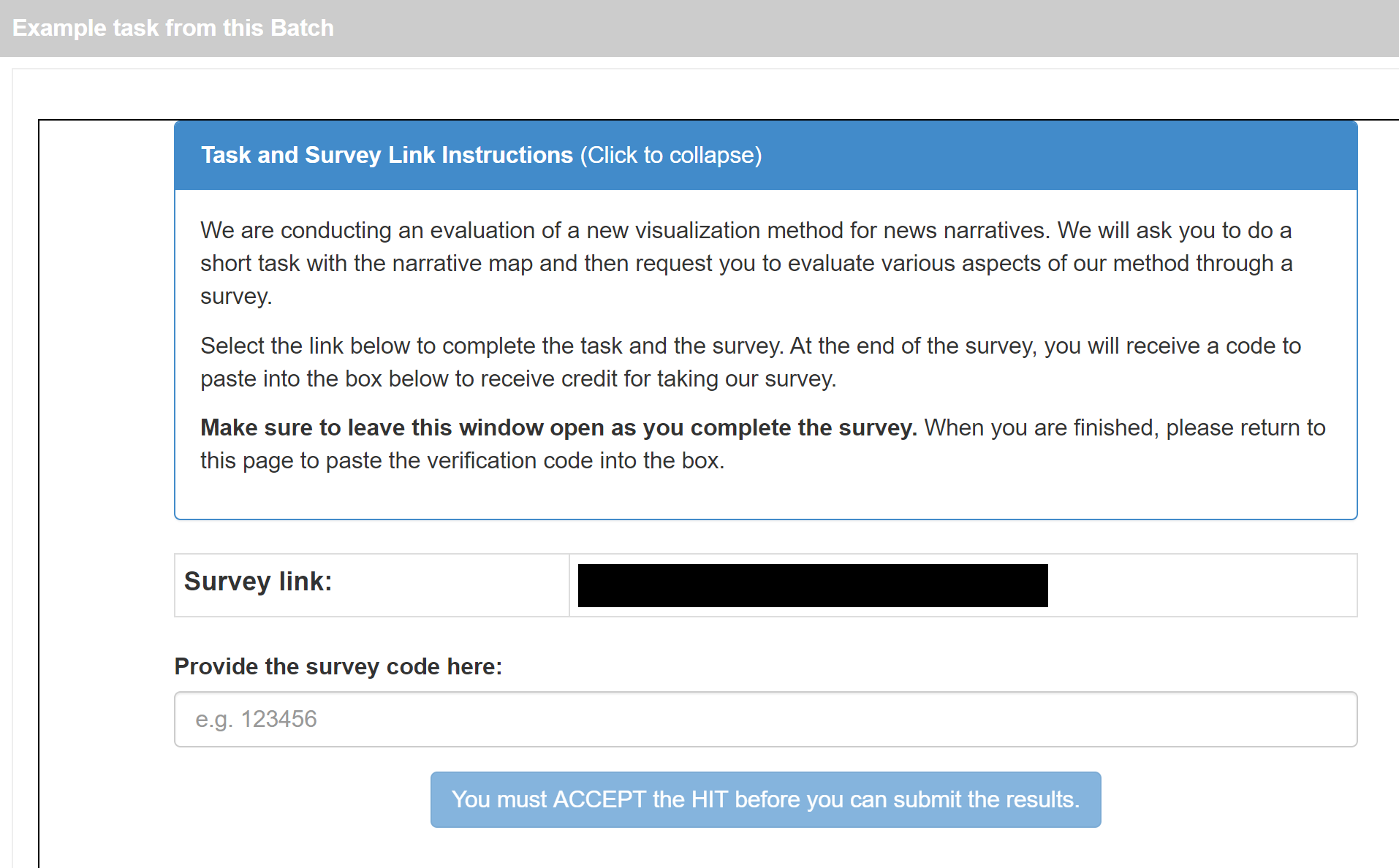}
    \caption{Description of the MTurk job shown to the workers.}
    \label{fig:MT2}
\end{figure}
\begin{figure}[htbp]
    \centering
    \includegraphics[width=0.7\textwidth]{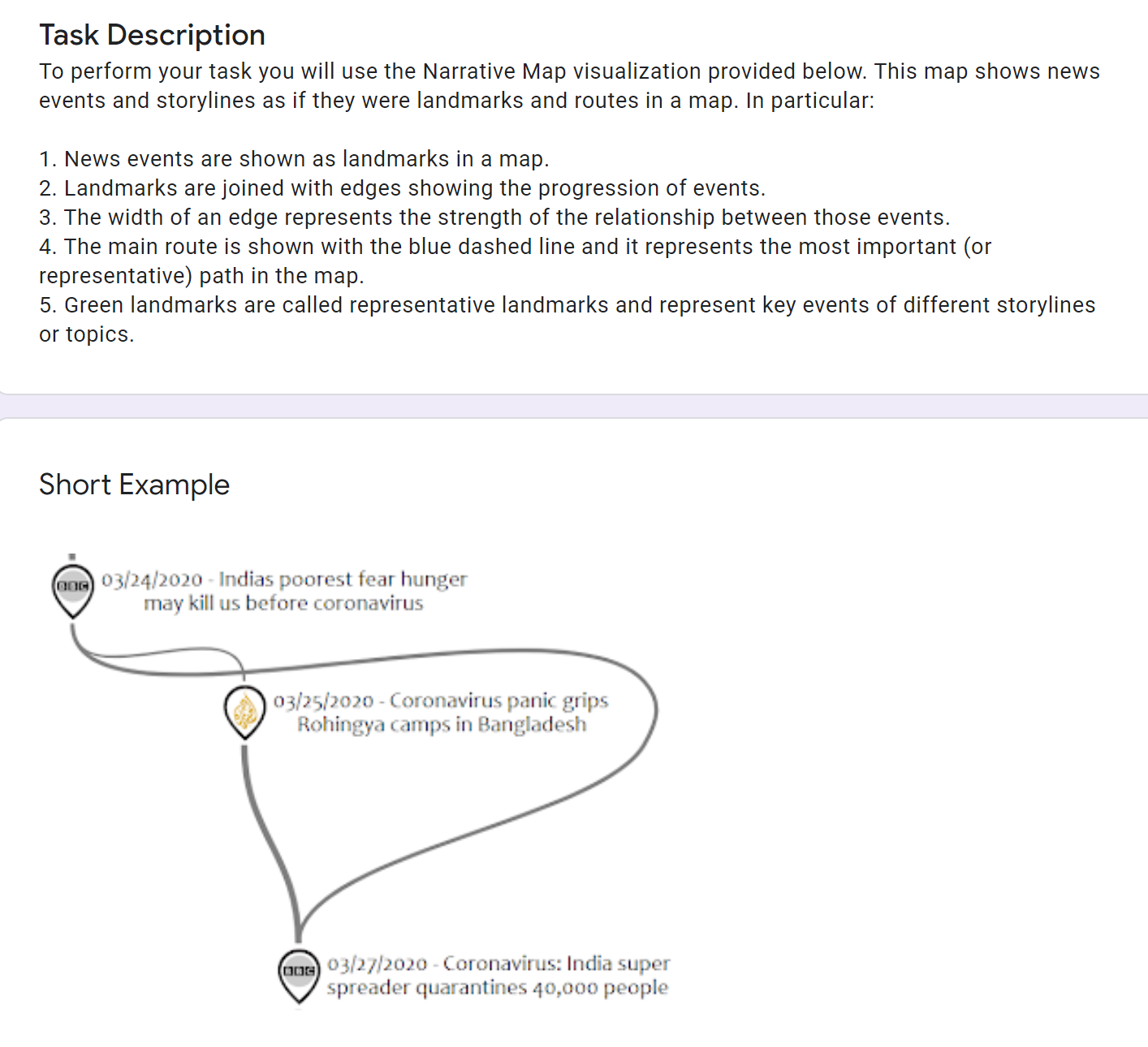}
    \caption{Narrative Map Description and Short Example}
    \label{fig:MT3}
\end{figure}
\begin{figure}[htbp]
    \centering
    \includegraphics[width=0.7\textwidth]{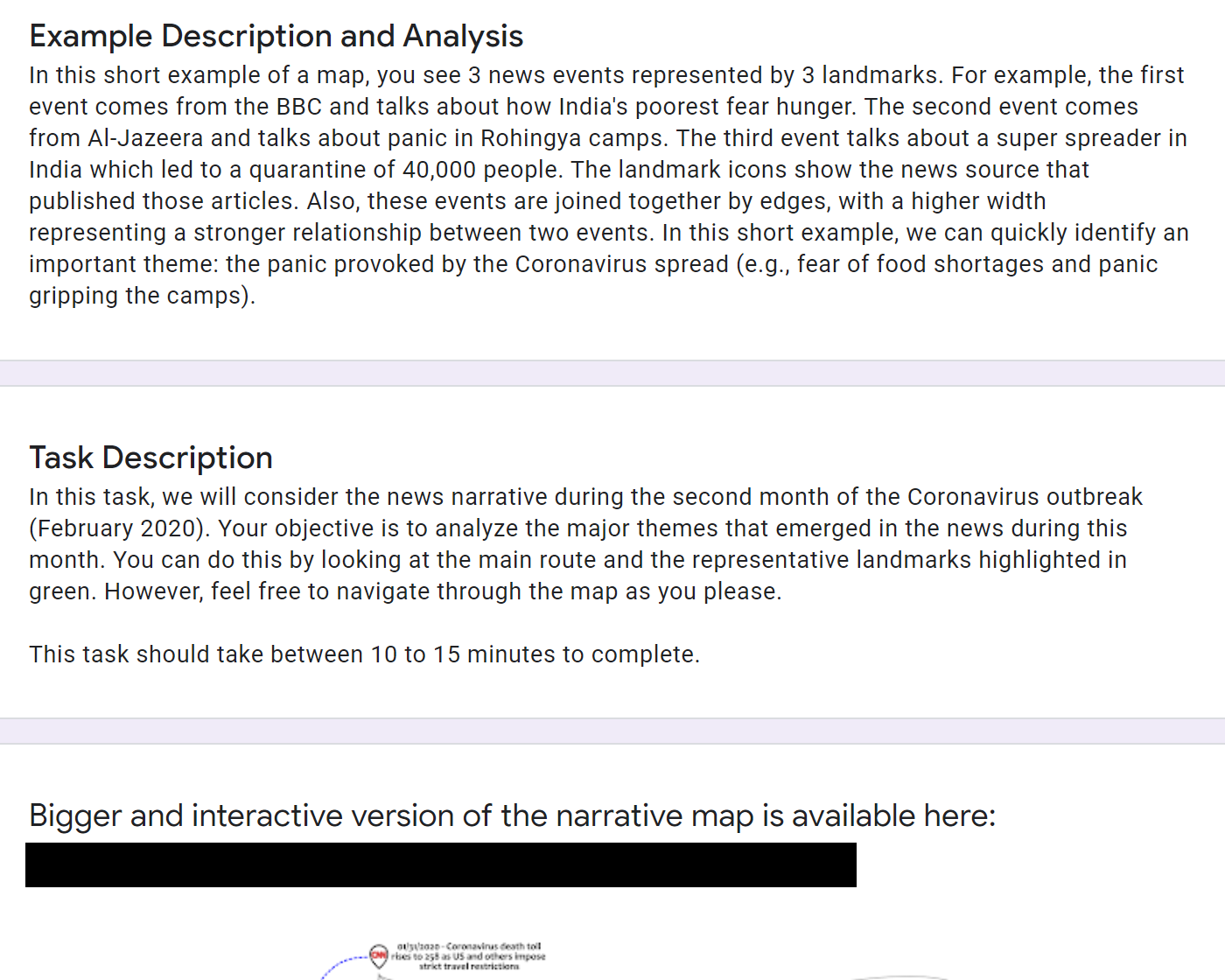}
    \caption{Example Analysis and Assigned Task}
    \label{fig:MT4}
\end{figure}
\begin{figure}[htbp]
    \centering
    \includegraphics[width=0.7\textwidth]{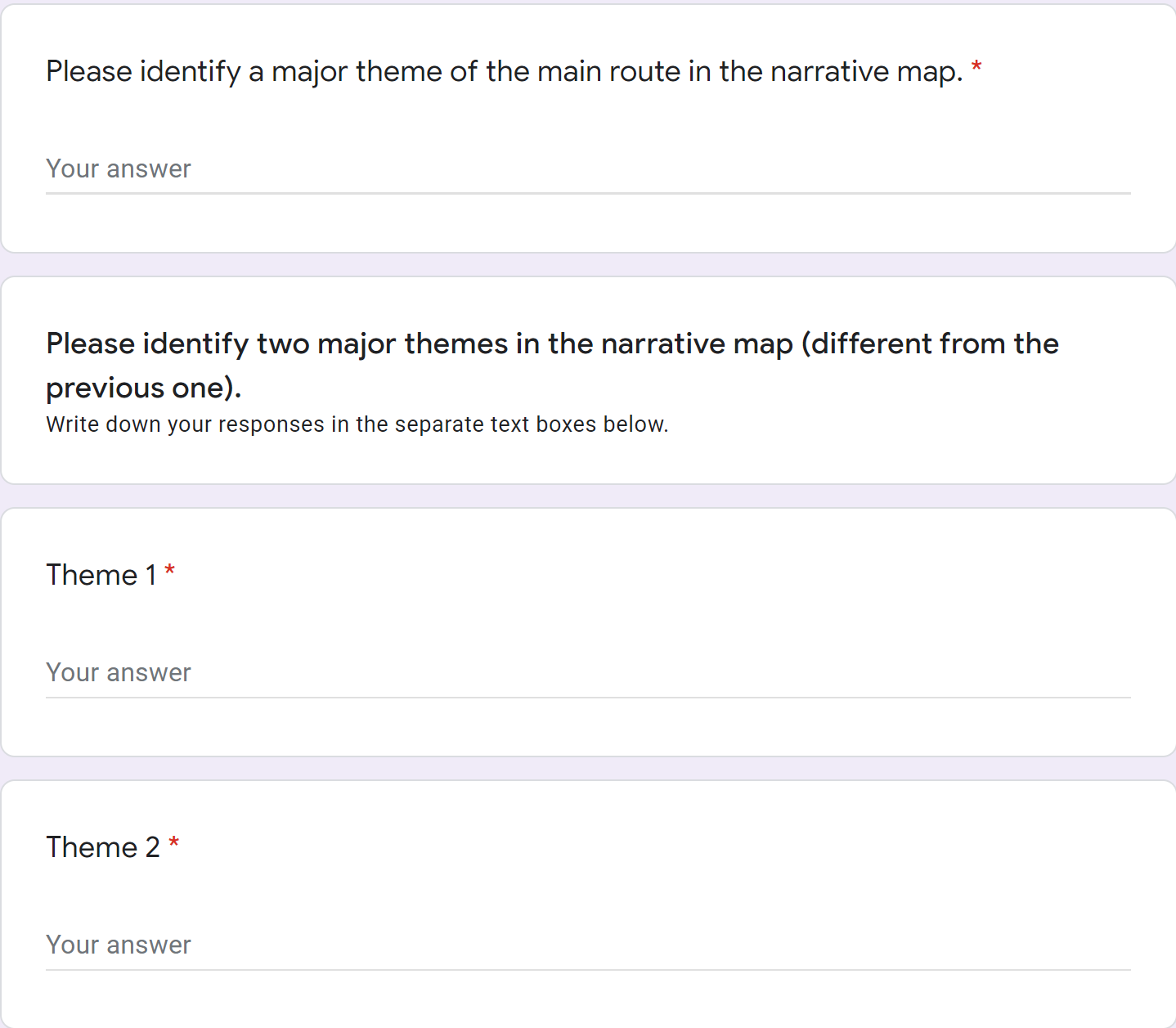}
    \caption{Theme identification task.}
    \label{fig:MT5}
\end{figure}

\end{document}